\documentstyle[aps,multicol,tighten,eqsecnum]{revtex} 
 
\begin{document} 
\draft 
\date{\today}
\title{Adaptive single-shot phase measurements: The full quantum theory} 
\author{H.M.~Wiseman$^{1,*}$ and R.B.~Killip$^{2}$} 
\address{$^{1}$Department of Physics, The University of Queensland, 
St.~Lucia 4072, Australia \\ 
$^{2}$Division of Physics, Mathematics, and Astronomy, 
California Institute of Technology, CA 91125, U.S.A.}
\maketitle
\begin{abstract}
	The phase of a single-mode field can be measured in a single-shot 
	measurement by interfering the field with 
	an effectively classical local oscillator of known phase. The standard 
	technique is to have the local oscillator detuned from the system 
	(heterodyne detection) so that 
	it is sometimes in phase and sometimes in quadrature with the system over 
	the course of the measurement. This enables both quadratures of the 
	system to be measured, from which the phase can be estimated. One of us 
	[H.M. Wiseman, Phys. Rev. Lett. {\bf 75}, 4587 (1995)]  has 
	shown recently that it is 
	possible to make a much better estimate of the phase by using an adaptive 
	technique in which a resonant local oscillator has its phase adjusted by 
	a feedback loop during the single-shot measurement. In Ref.~[H.M. Wiseman 
	and R.B. Killip, Phys. Rev. A {\bf 56}, 944] we presented a semiclassical 
	analysis of a particular adaptive 
	scheme, which yielded asymptotic results for the 
	phase variance of strong fields. In this paper 
	we present an exact quantum mechanical 
	treatment. This is necessary for calculating the phase variance for 
	fields with small photon numbers, and also for considering figures of 
	merit other than the phase variance. Our results show that an adaptive 
	scheme is always superior to heterodyne detection as far as the 
	variance is concerned. However the tails 
	of the probability distribution are surprisingly high for this adaptive 
	measurement, so that it does  not always result in a smaller 
	probability of error in phase-based optical communication. 
\end{abstract}

\pacs{42.50.Dv, 42.50.Lc}

\begin{multicols}{2}

\newcommand{\beq}{\begin{equation}} 
\newcommand{\eeq}{\end{equation}}
\newcommand{\bqa}{\begin{eqnarray}} 
\newcommand{\eqa}{\end{eqnarray}}
\newcommand{\erf}[1]{Eq.~(\ref{#1})}
\newcommand{\nn}{\nonumber} 
\newcommand{\dg}{^\dagger}
\newcommand{\smallfrac}[2]{\mbox{$\frac{#1}{#2}$}}
\newcommand{\bra}[1]{\langle{#1}|} 
\newcommand{\ket}[1]{|{#1}\rangle}
\newcommand{\ip}[1]{\left\langle{#1}\right\rangle}
\newcommand{\sch}{Schr\"odinger } 
\newcommand{\schs}{Schr\"odinger's }
\newcommand{\hei}{Heisenberg } 
\newcommand{\heis}{Heisenberg's }
\newcommand{\half}{\smallfrac{1}{2}} 
\newcommand{\bl}{{\bigl(}}
\newcommand{\br}{{\bigr)}} 
\newcommand{\ito}{It\^o }
\newcommand{\str}{Stratonovich } 
\newcommand{\hv}{\hat\varphi}
\newcommand{\implies}{\Longrightarrow}

\section{Introduction}

In a  typical textbook of 
quantum mechanics one might find a statement such as
\begin{quotation}
Every physical quantity ${\cal Z}$ has associated with it an 
Hermitian operator $Z$. A measurement of ${\cal Z}$ for a system with 
state matrix $\rho$ will yield a result 
$z$ which is an eigenvalue of $Z$. The probability of getting the 
result $z$ is equal to  $\bra{z}\rho\ket{z}$ where $Z\ket{z}=z\ket{z}$.
\end{quotation}
Unfortunately the number of measurements of physical quantities 
for which this quantum measurement theory applies is very small. 
Nevertheless there are some in the context of quantum optics. It is 
only detector inefficiencies (now quite small) which limit the 
measurement of the photon number with operator $a\dg a$ 
and quadratures with operators such 
as $X=a+a\dg$  
for single-mode optical fields. The former can be measured by direct 
photon counting and the latter by adding an essentially classical 
field of known phase (called the local oscillator) to the 
quantum field before counting photons (see for example 
Ref.~\cite{Wis95qo1}). 

There is one 
obvious optical quantity of which we cannot make a quantum-limited 
measurement: the phase $\phi$ of the electromagnetic field. Despite 
the difficulties in  defining a phase operator 
(which can be overcome \cite{PegBar89}),  the ``phase  
eigenstates'' $\ket{\phi}$ are independent of any phase operator (see 
Sec.~\ref{22}) and
have been recognized for a very long time \cite{Lon27}. The
opinion is sometimes expressed 
that the reason one cannot measure phase is that the phase 
eigenstates  do not have (even approximately) 
compact support on the number states, so that a measurement of phase 
would require infinite energy. This argument is specious, because the 
eigenstates of $a+a\dg$ also do not have compact support on the energy 
eigenstates, and yet in the limit 
of infinite local oscillator strength and perfect photodetection 
a homodyne 
measurement approaches a quadrature measurement. 
Nevertheless it is true that phase cannot 
be measured exactly, even in these ideal limits. The reason for this 
will be explored in the discussion section. 

Although the quantum phase of a single mode field cannot be measured 
exactly, it can be measured approximately. As well as being 
interesting for theoretical reasons, there may be 
 practical reasons for wishing to measure phase.
 For example, quantum-limited communication could 
be possible by encoding information in the phase of single-mode 
pulses of light. The first requirement for such a scheme would be 
to create states with very
well-defined phase. This has been investigated by various authors (see 
 Ref.~\cite{PSphase93} for some of these). The next step would be encoding the 
signal, which is easy to do using an
electro-optic modulator. The third requirement is for the receiver to measure the 
encoded phase as accurately as possible. This is a problem which 
seems not to have received the amount of attention it deserves, given 
that it is as important to communication as the generation of states 
with well-defined phase.
Another application for accurate phase measurements 
could be in inferring the properties of other quantum systems which 
can cause a phase shift, such as the presence of an atom at a 
particular point in a single-mode standing wave. 

The standard way of measuring phase (approximately) is to use two 
simultaneous homodyne measurements of orthogonal quadratures (known as
eight-port homodyne detection), or heterodyne detection, which are 
equivalent in an appropriate limit \cite{LeoVacBohPau95}. 
A way to improve upon this was 
first suggested by one of us \cite{Wis95c}: single-shot 
adaptive measurements. 
By this we mean the use of measurement results from earlier stages of 
a {\em single measurement} to affect the conditions of the 
measurement in its later stages. In this case it means using the 
photocurrent up to time $t$ to control the local oscillator phase at 
time $t$ by a feedback loop, during the detection of a single 
single-mode pulse. In 
Ref.~\cite{WisKil97} we investigated a particular feedback algorithm, 
illustrated in Fig.~1, 
using semiclassical theory. We showed that for large fields an adaptive 
measurement is a much closer approximation to a true phase measurement 
than is heterodyne detection. 

In this paper we continue our analysis of the simple adaptive 
algorithm, but this time we present the full quantum theory of these adaptive 
phase measurements. The 
background theory required is presented in Sec.~II. This introduces 
the theory of probability-operator-measures (POMs) 
which is required for approximate measurements. It also summarizes 
the theory of POMs for phase measurements and POMs for measurements 
using a large local oscillator. In Sec.~III we derive
expressions for the POMs for the two adaptive phase measurement 
schemes of 
Ref.~\cite{WisKil97}. In Sec.~IV we use these POMs to calculate 
phase variances, for coherent states and for phase-optimized states with 
an upper bound on the photon number. 
We compare our exact (quantum) numerical results to the 
asymptotic (semiclassical) analytical results obtained in 
Ref.~\cite{WisKil97}.
One feature which can only be calculated using the full quantum theory 
is the overall shape of the probability distributions, including the 
tails. This is required for determining the 
probability of error in phase communication schemes. This aspect is 
investigated in Sec.~V, again for coherent states and for 
phase-optimized states with 
an upper bound on the photon number. Sec.~VI concludes with a 
discussion on the ultimate limits to phase measurements.

\section{Probability-Operator-Measures}

\subsection{General theory of POMs}

If (as in the present case) 
we are unconcerned about the fate of the system after it has 
been measured, then any measurement is completely described by the 
probability for each of the possible results to occur. Let 
the set of all possible measurement results $\lambda$ be denoted $\Omega$. 
Then the 
measurement is specified by a probability-measure (PM)  on 
$\Omega$. If we denote the PM as $P$ then for any subset $E \subseteq 
\Omega$, we 
can identify $P(E)$ as the probability to obtain a measurement result 
$\lambda \in E$. Of course this requires $P(\Omega)=1$. 

For quantum mechanical systems, the most general way of generating a 
PM $P$ is as the expectation value of an operator-measure $F$ 
on $\Omega$. 
That is, for a quantum system with state matrix $\rho$,
\beq
P(E) = {\rm Tr}[\rho F(E)].
\eeq
Obviously $F(E)$ must be a positive operator, and by conservation of 
probability
\beq \label{complet1}
F(\Omega) = 1.
\eeq
For this reason we call $F$ a Probability-operator-measure 
(POM), or sometimes an {\em effect}-valued measure \cite{Dav76,Hel76}. 
Note that even for a subset $E$ with a single element $\lambda$, 
$F(\lambda)$ is not necessarily a projector.

\subsection{POMs for phase measurements}

\label{22}
Now consider the case where the measured quantity is to be a phase 
$\phi$ of a single-mode photon field, 
so that $F$ is a POM on $\Omega = [0,2\pi)$. 
Quantum 
mechanically this phase should in some sense be conjugate to the photon
number operator $a\dg a$, but as long as we stick with POMs to 
describe the measurement there are none of the difficulties associated 
with defining a phase operator \cite{PegBar89}.
Since phase is a continuous variable, we will use $F(\phi)$ to denote 
the phase POM density. 
The completeness relation for a phase POM is therefore written as
\beq \label{complet2}
\int_{0}^{2\pi} d\phi F(\phi) = 1.
\eeq
As explained in Ref.~\cite{WisKil97}, for $F(\phi)$ to be invariant 
under phase shifts, and to be unbiased, implies that it can be 
written in the form
\beq \label{genPOVM}
F(\phi) = \frac{1}{2\pi} \sum_{n,m=0}^{\infty} \ket{m}\bra{n} 
e^{i\phi(m-n)} H_{mn}.
\eeq
Here $H$ is a positive-semidefinite Hermitian matrix with all entries real and 
positive, and $\ket{m}$ is the number state $a\dg a \ket{m} = m\ket{m}$.

The completeness condition (\ref{complet2}) implies that
\beq
\forall\; m\geq 0 \;\; H_{mm}=1.
\eeq
The positivity condition on the matrix $H$ obviously requires that the 
off-diagonal elements be less than or equal to unity. A unique phase 
measurement is defined by specifying that all of the off diagonal 
elements be equal to unity. This is what has recently been 
called a 
canonical phase measurement \cite{LeoVacBohPau95}, although its 
special role was 
recognized very early in the history of quantum theory \cite{Lon27}. 

In realistic phase 
measurements the off-diagonal elements $H_{m,n}$ will be less than unity, 
but for $|m-n|=1$ and $m \gg 1$ they should be close to unity if 
the measurement is to be a good phase measurement, as will be seen 
in Sec.~IV. In fact, in all of the measurements we examine, we have 
\beq \label{defh}
h(m) \equiv 1 - H_{m,m+1} \leq O(m^{-1/2}).
\eeq
For a canonical measurement $h(m)$ is identically zero. In this case 
we can write the POM (\ref{genPOVM}) as
\beq
F^{\rm can}(\phi) =  \frac{1}{2\pi} \ket\phi \bra\phi,
\eeq
where $\ket\phi$ is an unnormalized phase eigenstate
\beq
\ket\phi = \sum_{n=0}^{\infty} e^{in\phi}\ket{n}
\eeq
as referred to in the introduction.

\subsection{POMs for dyne measurements}

We now turn from the POMs for phase measurements of a single-mode 
field to the POMs for 
measurements on a single-mode photon field made by interfering the 
light from that field with another field which has a macroscopic 
coherent excitation. This can be done at a beam splitter, and the two 
output fields of the beam splitter can then be detected by normal 
photodetectors. The second field can be treated classically as 
a $c$-number, and is known as a local oscillator. All practical phase-sensitive 
measurements require a local oscillator, to act as a 
phase reference. If the local 
oscillator is resonant with the system field then this type of 
measurement is known as {\em homodyne} detection. If the local oscillator 
is detuned (outside 
the bandwidth of the system field) then this is known as {\em 
heterodyne} detection. In considering phase measurements we will have 
to consider other sorts of measurements involving interference with a 
quasiclassical local oscillator. In ignorance of any 
received term for such measurements we will call them examples of {\em dyne} 
detection, so that homodyne and heterodyne are obviously special cases.

Let us assume that our single-mode signal field has a temporal pulse-shape 
$u(t)$ which is positive and normalized as
\beq
\int_{0}^{T} u(t) = 1.
\eeq
Here we are obviously ignoring the phase variation at optical 
frequency $\omega$; $u(t)$ is the envelope function.
The total time $T$ is necessarily much greater than 
$\omega^{-1}$, so that the pulse can be considered monochromatic. 
This is essential in order for the dyne measurements (which are phase-sensitive 
measurements) to be quantum-limited. That is, for quantum effects 
to provide the limit to the phase uncertainty in the measurement. 
If the characteristic spectral width of the 
pulse $\Gamma \agt T^{-1}$ 
is too large then the phase uncertainty 
will be dominated by the term $\delta \phi \sim \Gamma/\omega$ coming 
from the uncertainty $\Gamma$ in the frequency. 
In all that follows we assume this 
uncertainty to be negligible. 

For simplicity we will take the beam splitter at which the system and local 
oscillator fields are interfered to be balanced (50/50). Then, 
ignoring vacuum fluctuations, the two 
fields at the two output ports of the beamsplitter are equal to
\beq
b_{\pm}(t) = \sqrt{u(t)/2} \left( a \pm \beta e^{i\Phi(t)} \right) 
e^{-i\omega t},
\eeq
where $a$ is the annihilation operator for the system and the real 
number $\beta$ is the coherent amplitude of the local oscillator. This 
is normalized so that the instantaneous 
rate of photodetection at each detector is 
$\langle b_{\pm}\dg(t) b_{\pm}(t) \rangle$. 
We have assumed that the intensity-profile of the local 
oscillator is the same as that of the system. However, we have 
included an arbitrary phase variation $\Phi(t)$ of the local 
oscillator relative to the system. The total number of photons in the 
local oscillator is $\beta^{2}$, so we are interested in the limit 
$\beta^{2}\gg 1, \langle a\dg a \rangle$. For homodyne detection 
$\Phi(t) = \Phi_{0}$, a constant. For heterodyne detection $\Phi(t) = 
\Phi_{0}+ t\Delta$, where $\Delta \gg \Gamma$ is the detuning.

The signal 
of interest is simply the difference between the two photocurrents at 
the two detectors (labeled $\pm$). 
If we denote the number of photocounts at each of  the  detectors 
 in the time interval $[t,t+\delta)$ by $\delta N_{\pm}(t)$ then 
 we can define the signal photocurrent as
\beq
I(t) = \lim_{\delta t \to 0}\; \lim_{\beta\to\infty} \frac{\delta N_{+}
(t) - \delta 
N_{-}(t)}{\beta \delta t} \\
\eeq
Note that the two limits here do not commute. The limit $\beta \to 
\infty$ implies that both photocounts will be dominated by the 
contribution from the local oscillator. The fact that the limit 
$\delta t \to 0$ is taken second 
indicates that we are only interested in the fluctuations in $I(t)$ on 
a time scale much greater than the mean time $\sim u(t)^{-1}\beta^{-2}$ 
between photodetections. 

The general quantum theory of dyne measurements was derived by one of 
us in Ref.~\cite{Wis96a} for the case where the system mode is 
derived from an exponentially decaying cavity so that $u(t) = 
\gamma e^{-\gamma t}$ where $\gamma$ is the cavity linewidth. This is easily 
generalized for arbitrary $u(t)$. First we define a scaled time 
variable
\beq
v = \int_{0}^{t} u(s)ds.
\eeq
This is dimensionless, and increases monotonically with $t$ from $0$ 
to $1$.
For the case $u(t)=\gamma e^{-\gamma t}$ we have $v = 1-e^{-\gamma t}$.
The photocurrent in terms of $v$ is scaled so that 
\beq
I(v) dv = I(t) dt = dv\,I(t)/u(t).
\eeq

Now the measurement result for a dyne measurement up to time $t$ 
is the complete photocurrent record $I(t')$ from $t'=0$ to $t'=t$ 
[or equivalently, $I(v')$ from 
$v'=0$ to $v'=v$)]. This record is, in theory at least, a continuous infinity 
of real numbers, which is an impractically huge amount of data. Fortunately it turns out 
that there are just two sufficient statistics at scaled time $v$ (henceforth 
called simply time), namely the two complex numbers
\bqa
A_{v} &=& \int_{0}^{v} I(u) e^{i\Phi(u)} du \label{defA} \\
B_{v} &=& -\int_{0}^{v} e^{2i\Phi(u)}du. \label{defB}
\eqa
We call these the sufficient statistics because, as shown in 
Ref.~\cite{Wis96a}, the POM for the measurement at time $0 \leq v < 1$ is given 
by
\bqa
G_{v}(A_{v},B_{v}) &=& Q_{v}(A_{v},B_{v}) 
\exp\bigl( \mbox{$\frac12$} B_v {a^\dagger}^2  + A_v
a^\dagger \bigr) \nn \\
&& \times (1-v)^{a^\dagger \! a/2} \exp \bigl( \mbox{$\frac12$} 
B^*_v a^2  + A^*_v a
\bigr), \label{foradapt}
\eqa
where $Q_{v} \bigl( A_v,B_v \bigr)$ is a positive function which will 
be defined shortly. This implies that the probability for obtaining 
any photocurrent $\{I(u):0\leq u<v\}$ is determined 
only by the two complex functionals of this current $A_{v}$ and 
$B_{v}$. Any other features of $\{I(u):0\leq u<v\}$ are completely irrelevant.

It might be thought that the second integral $B_{v}$ does not 
depend on $\{I(u):0\leq u<v\}$ at all, because the 
photocurrent does not appear explicitly in Eq.~(\ref{defB}). However, 
it may appear implicitly if the local oscillator phase $\Phi(v)$ 
depends upon $\{I(u):0\leq u<v\}$. This is precisely the situation we 
will consider later to construct a phase measurement. When we do so, 
the theory presented here shows that $\Phi(v)$ should be made to 
depend on $\{I(u):0\leq u<v\}$ only through the two integrals 
(\ref{defA}),(\ref{defB}). That is to say, we should have
\beq
\Phi(v) = f_{v}(A_{v},B_{v})
\eeq
for some (possibly time-dependent) function $f$. This is an extremely 
powerful result which is not at all intuitive.

In the limit $v \to 1$, $(1-v)^{a^\dagger \! a/2} \to 
\ket{0}\bra{0}$, where $\ket{0}$ is the vacuum state. So, dropping 
the subscript $v$ when $v=1$,  we can 
write the POM (\ref{foradapt}) as
\beq \label{FAB2}
G(A,B) = Q(A,B) \ket{\tilde{\psi}(A,B)}\bra{\tilde{\psi}(A,B)},
\eeq
where $\ket{\tilde{\psi}(A,B)}$ is an unnormalized 
ket defined by
\beq \label{deftildepsi}
\ket{\tilde{\psi}(A,B)} = \exp\bigl( \mbox{$\frac12$} B {a^\dagger}^2  + A
a^\dagger \bigr) \ket{0}.
\eeq
With a little operator algebra it is easy to show that this is 
proportional to the squeezed state \cite{WalMil94}
\beq \label{ss0}
\ket{\alpha, \epsilon} = \exp(\alpha a\dg - \alpha^{*} a)
\exp\bigl( \mbox{$\frac12$} \epsilon^{*} a^{2} - \epsilon {a\dg}^{2} 
\bigr) \ket{0},
\eeq
where
\bqa
\alpha &=& \frac{A+BA^{*}}{1-|B|^{2}} \label{ss1} \\
\epsilon &=& \frac{ - B\, {\rm atanh}|B|}{|B|}. \label{ss2}
\eqa
From Eq.~(\ref{defB}), it is evident that $|B| \leq 1$. For the 
schemes we will consider $|B|< 1$ with probability one, so 
that the two expressions (\ref{ss1}), (\ref{ss2}) 
are well-defined. 

If we rewrite the POM (\ref{foradapt}) in terms of $\alpha,\epsilon$ 
instead of $A,B$, we have 
\beq
G'(\alpha,\epsilon) = Q'(\alpha,\epsilon) 
\ket{\alpha,\epsilon}\bra{\alpha,\epsilon},
\eeq
where $Q'$ is some new positive function of $\alpha,\epsilon$. In this 
case the set of all measurement results is $\Omega={\bf C}\otimes {\bf C}$, 
where ${\bf C}$ denotes the set of complex numbers. If we imagine 
varying the state of the system $\ket{\psi}$ (assumed pure), then the 
probability to obtain the result $\alpha,\epsilon$ is
\beq \label{weird}
P(\alpha,\epsilon) \propto |\ip{\alpha,\epsilon | \psi}|^{2} .
\eeq
Provided $\exp(|\epsilon|) \ll |\alpha|$, the squeezed state 
$\ket{\alpha,\epsilon}$ has a well-defined coherent amplitude $\alpha$.
Hence from Eq.~(\ref{weird}) if the unknown 
 system state $\ket{\psi}$ is also localized in the 
phase plane, it is highly likely that it must have a coherent 
amplitude close to $\alpha$. This fact will be used later to good 
effect.

We must now address the issue of how $Q(A,B)$ is found. In Ref.~\cite{Wis96a} 
it is shown that  $Q(A,B)$ is the 
joint probability distribution that $A,B$ would have if the 
photocurrent $I(v)$ were given by
\beq \label{ostcur}
I(v)dv = dW(v)
\eeq
where $dW(v)$ is the infinitesimal increment in a real Wiener 
process \cite{Gar85} satisfying 
\bqa
\ip{dW(v)} &=& 0,\label{meandW} \\
dW(v) dW(v) &=& dt. \label{dW2dt}
\eqa 
In Ref.~\cite{Wis96a}, $Q(A,B)$ was called the {\em 
ostensible} probability distribution for $A,B$. It is the probability 
distribution that $A,B$ would have if there were no signal whatsoever; 
that is, if the system were prepared in the vacuum state. The noise in 
Eq.~(\ref{ostcur}) then represents the local oscillator shot noise (or 
vacuum fluctuations if a Heisenberg picture interpretation 
is preferred). The presence of a non-zero signal determines the {\em 
actual} probability distribution through the POM (\ref{FAB2}). That 
is to say, if the system state matrix is $\rho$ then the true probability 
density is
\beq
P(A,B) d^{2}A d^{2}B = Q(A,B) \bra{\tilde{\psi}(A,B)}\rho 
\ket{\tilde{\psi}(A,B)}d^{2}A d^{2}B.
\eeq

Before moving onto specific examples in the following section, 
we will derive some general results regarding the ostensible 
distribution $Q(A,B)$. First, the ostensible mean of $A$ is
\beq \label{A1}
\ip{A}_{Q} = \int_{0}^{1} \langle e^{i\Phi(v)} dW(v) \rangle = 
0.
\eeq
This holds true even if $\Phi(v)$ depends on the photocurrent record 
$\{I(u): 0\leq u <v\}$ because $W(v)$ is a strictly Markovian process.
Second, 
\bqa \label{A2}
\ip{A^{2}}_{Q} &=& \int_{0}^{1}  \int_{0}^{1} \langle e^{i\Phi(v)+i\Phi(u)}
 dW(u) dW(v)\rangle \nn
 \\ &=& \int_{0}^{1} dv \ip{e^{2i\Phi(v)}} = 
 -\langle B \rangle_{Q}.
\eqa
Third,
\beq \label{A3}
\ip{|A|^{2}}_{Q} = \int_{0}^{1} \int_{0}^{1} \ip{dW(u)dW(v)} = 
\int_{0}^{1} dv = 1.
\eeq

\section{Physically Realizable Phase Measurements}

\subsection{Heterodyne Measurements}

As noted in Sec.~\ref{22} the ideal form of phase measurement is a 
canonical phase measurement in which $H_{mn}$ from Eq.~(\ref{genPOVM}) is 
equal to unity for all $m,n$. This is plotted in Fig.~2(a). 
All physically realizable 
phase measurements fall short of this ideal. The simplest method for 
making a phase measurement is via heterodyne detection. As explained 
above, this involves a local oscillator which is far detuned from the 
system. The linear variation of the phase is in fact not essential; 
all that is required is that all relative phases (of the system with 
respect to the local oscillator) be sampled equally and on a time scale 
much shorter than the reciprocal bandwidth of the system. As long as 
there is a record of the local oscillator phase as a function of time, 
the information in the photocurrent record can be recovered. For 
definiteness, however, we will take the local oscillator phase to 
simply change linearly with (scaled) time $v$. That is,
\beq
\Phi(v) = \Phi_{0} + v \Delta,
\eeq
where $\Delta \gg 1$. 

Having specified $\Phi(v)$ all that remains to completely describe 
this heterodyne measurement is to determine $Q(A,B)$, the ostensible 
probability distribution for the measurement results $A,B$. Because 
the above $\Phi(v)$ is independent of the photocurrent $I$, the 
`result' $B$ is 
a constant (rather than a random variable) with value
\bqa
B &=& -\int_{0}^{1} dv \exp[2i(\Phi_{0}+v\Delta)] 
\\ &= &
{\exp(2i\Phi_{0})}\frac{1-\exp(2i\Delta)}{2i\Delta} \to 0,
\eqa
where the final limit results from taking $\Delta \to \infty$.
The only variable in this case is therefore 
\beq
A = \int_{0}^{1} dv I(v) \exp[i(\Phi_{0}+v\Delta)].
\eeq
To find the ostensible statistics for $A$ we treat $I(v)dv$ as an independent 
Gaussian variable $dW(v)$ for each infinitesimal interval $[v,v+dv)$. 
Since $A$ is just the sum of these Gaussian 
variables, it must ostensibly be a Gaussian variable itself. 
From Eqs.~(\ref{A1})--(\ref{A3}) with $B=0$ it follows that the ostensible 
distribution for $A$ is the rotationally-invariant Gaussian 
\beq \label{hetost}
Q^{\rm het}(A)d^{2}A = \pi^{-1} \exp(-|A|^{2}) d^{2}A.
\eeq

From these results and Eq.~(\ref{FAB2}) we find the POM for 
heterodyne measurements to be
\beq \label{hetPOVM1}
G^{\rm het}(A) = \pi^{-1} \exp(-|A|^{2}) \ket{\tilde{\psi}(A,0)}
\bra{\tilde{\psi}(A,0)}.
\eeq
Now from Eqs.~(\ref{ss0})--(\ref{ss2}) it is easy to verify that 
$\ket{\tilde{\psi}(A,0)}$ is simply proportional to the coherent 
state $\ket{A}$ where $A$ is the coherent amplitude usually denoted $\alpha$. 
It turns out that the 
proportionality factor is just $\exp(|A|^{2}/2)$ so that we can 
rewrite \erf{hetPOVM1} as 
\beq \label{hetPOVM2}
G^{\rm het}(A) = \pi^{-1} \ket{A}\bra{A}
\eeq
This result has been obtained many times before by other means; 
for one example see Ref.~\cite{Wis95qo1}. The factor of $\pi^{-1}$ 
remains because the coherent states are overcomplete.

In the context of this paper we are interested in heterodyne 
measurements only in so far as they enable us to make an estimate 
of the phase of the system. If there is no prior information about 
the system then Eq.~(\ref{hetPOVM2}) suggests a good estimate 
of the phase to be
\beq \label{phihet}
\phi_{\rm het} = \arg A.
\eeq
The POM for this phase estimate is found simply by marginalizing the 
modulus of $A$. That is,
\beq \label{hetPOVM3}
F^{\rm het}(\phi) = \int_{0}^{\infty} |A|\, d(|A|) \, G^{\rm 
het}(|A|e^{i\phi}) 
\eeq
Evaluating this in the number state basis yields the matrix $H$ of 
\erf{genPOVM} to be
\beq
\label{heterH}
H^{\rm het}_{mn} = \frac{\Gamma \left( \frac{n+m}{2}+1 \right)}{\sqrt{n!m!}}.
\eeq
Clearly $H^{\rm het}_{nn}=1$ as required, while the off-diagonal elements 
decrease with distance away from the diagonal. These features can be 
seen in the matrix plot of $H^{\rm het}_{mn}$ in Fig.~2(b).

\subsection{Adaptive Measurements}

A heterodyne phase measurement is not as good as a canonical 
measurement because it is actually a measurement of both phase and 
amplitude, with the latter information being thrown away. In order to 
make a better phase measurement one would like to concentrate on 
measuring the phase quadrature. This can be done by homodyne 
detection \cite{WisKil97}, but only if one already knows the 
phase of the system. A true phase measurement should work even if 
one has no information about the system phase. Nevertheless we can 
use this idea to construct a true phase measurement as follows. 
Rather than measuring a fixed quadrature, we control the local 
oscillator phase as a function of time in order to measure the {\em 
estimated} phase quadrature. That is, we set $\Phi(v)$ to be equal to
\beq \label{fa1}
\Phi(v) = \hv(v) + \pi/2,
\eeq
where $\hv(v)$ is the {\em estimated} phase of the system at time $v$. 

Two questions remain to be decided. First, given our measurement record
$\{I(u):0\leq u <v\}$ how do we decide $\hv(v)$?
Second, what do we
choose to be our best estimate of phase $\phi$ 
once the measurement is completed?
We will postpone answering the second question. It was 
already noted above that the theory of dyne measurements implies that 
we should choose $\hv(v) = f_{v}(A_{v},B_{v})$  for some function $f$. 
For the remainder of this paper we choose 
\beq \label{fa2}
\hv(v) = \arg A_{v},
\eeq
as in Ref.~\cite{WisKil97}. As outlined in that reference, 
the motivations for this choice are:
\begin{enumerate}
	\item It is suggested by the above analysis for heterodyne 
	detection.

	\item  As shown by one of us \cite{Wis95c}, it reproduces the 
	canonical result if the system has at most one photon.

	\item  It gives a feedback algorithm 
	 which would be easy to implement experimentally.
	
	\item  It is mathematically tractable.
\end{enumerate}

When we say it can be exactly solved, we mean that we can 
determine the POM. To do this requires only the ostensible probability 
distribution $Q^{\rm ad}(A,B)$ given the feedback algorithm 
\erf{fa1}--(\ref{fa2}). To find this it is convenient to recast the 
ostensible integral equations (\ref{defA}),(\ref{defB}) as the 
ostensible \ito stochastic differential equations
\bqa
dA_{v} &=& e^{i\Phi(v)} dW(v), \\
dB_{v} &=& e^{2i\Phi(v)} dv,
\eqa
with the initial conditions
\beq
A_{0} = B_{0} = 0.
\eeq

With the above feedback algorithm we have $e^{i\Phi(v)} = 
iA_{v}/|A_{v}|$. This gives
\beq
dA_{v} = i A_{v} dW(v) / |A_{v}|.
\eeq
This can be solved by transforming to polar co-ordinates $\hv(v) = 
\arg A_{v}$ and $|A_{v}|^{2}$. Using the \ito calculus we find
\bqa
d |A_{v}|^{2} &=& dv ,\\
d\hv(v) &=& dW(v) / |A_{v}|.
\eqa
The first of these can be solved trivially to yield $|A_{v}| = 
\sqrt{v}$. That is, the modulus of $A$ evolves deterministically and in 
particular $|A| = 1$, as required by \erf{A3} 
Substituting this into the second gives
\beq \label{solnhv}
\hv(v) = \hv(0) + \int_{0}^{1} dW(v)/\sqrt{v}.
\eeq
Here $\hv(0)$ is an arbitrary initial phase. It is irrelevant to the 
problem because the divergence at $v=0$ of the integrand in this equation 
means that the initial phase will be randomized immediately:
\beq
\ip{\hv^{2}}_{Q} = \int_{0}^{1} dv/v = \infty.
\eeq
Thus the ostensible probability distribution for $A$ is
\beq \label{ostA}
Q_{a}^{\rm ad}(A) d^{2}A = \delta(|A|-1) |A| d(|A|) \frac{1}{2\pi} d(\arg A).  
\eeq

We require the joint ostensible probability distribution $Q^{\rm ad}(A,B)$. 
But rather than work with $B_{v}$ it is more convenient to consider 
the variable
\beq \label{intC}
C_{v} = e^{-2i\hv(v)}\int_{0}^{v} e^{2i\hv(u)} du.
\eeq
It is easy to prove that for $v=1$ 
\beq
C = B A^{*}/A,
\eeq
so that $A,C$ can replace $A,B$ as the sufficient statistics.
The advantage of the variable $C_{v}$ is that, from \erf{intC} and 
\erf{solnhv}, it obeys 
the stochastic \ito differential equation
\beq \label{deC}
dC_{v} = - \left[ \frac{2idW(v)}{\sqrt{v}} + \frac{2dv}{v}\right] 
C_{v}+ dv,
\eeq
with the initial condition $C_{0} = 0$. Since neither this initial 
conditions nor the above differential equation involve the value of 
$\hv(0)$ (which is essentially random as noted above), the final value 
of $C$ will be ostensibly independent of that of $A$. That is,
\beq
Q^{\rm ad}(A,C)=Q^{\rm ad}_{a}(A)Q^{\rm ad}_{c}(C).
\eeq
In fact, given the above result \erf{ostA} we need only $\hv = \arg A$ 
so that 
\beq
Q^{\rm ad}(A,C)d^{2}A\, d^{2}C \to \frac{d\hv}{2\pi} Q^{\rm ad}_{c}(C) d^{2}C.
\eeq

The problem remaining is thus to find $Q^{\rm ad}_{c}(C)$. It has not proven 
possible to find this analytically. However we have been able to find 
the exact values of the moments 
\beq
M^{n,m}_{v} = \langle C_{v}^{n} {C_{v}^{*}}^{m} \rangle_{Q}
\eeq
via a recurrence relation. This is done in Appendix A. For our 
purposes these moments are sufficient so we can assume the 
distribution $Q_{c}(C)$ known. From \erf{FAB2} The POM for the 
results $\hv,C$ under the feedback algorithm (\ref{fa1})--(\ref{fa2}) is 
thus
\bqa 
G^{\rm ad}(\hv,C)\, d\hv\, d^{2}C &=&\ket{\tilde{\psi}(e^{i\hv},e^{2i\hv}C)}
\bra{\tilde{\psi}(e^{i\hv},e^{2i\hv}C)}
 \nn \\
&& \times\, \frac{d\hv}{2\pi} d^{2}C Q^{\rm ad}_{c}(C).
\label{adPOVM1}
\eqa

Since the point of this exercise is to construct a phase measurement, 
we want ultimately to calculate some phase $\phi_{\rm ad}(\hv,C)$ 
from the sufficient 
statistics $\hv,C$. We are not constrained to choose $\hv$ even 
though we have been using it as our estimated phase in the feedback 
loop. Therefore the general expression for the POM of our adaptive phase 
measurement is
\beq
F^{\rm ad}(\phi)  =\int_{0}^{2\pi} d\hv \int\!\!\! \int d^{2}C 
\,G^{\rm ad}(\hv,C) 
\delta \bl 
\phi - \phi_{\rm ad}(\hv,C) \br.
\eeq

There are constraints on the function $\phi_{\rm ad}(\hv,C)$. Clearly 
if the phase of the state $\rho$ is rotated by some angle $\theta$, the 
probability distribution $P_{\rm ad}(\phi) = {\rm Tr}[\rho 
F_{\rm ad}(\phi)]$ for $\phi$ should be shifted similarly. Now to rotate the 
phase of the state by $\theta$ is equivalent to rotating that of the 
POM by $-\theta$. This has the effect of replacing 
$\ket{\tilde{\psi}(e^{i\hv},e^{2i\hv}C)}$ by
\beq
e^{-i\theta a\dg a} \ket{\tilde{\psi}(e^{i\hv},e^{2i\hv}C)}
= \ket{\tilde{\psi}(e^{i(\hv-\theta)},e^{2i(\hv-\theta)}C)}
\eeq
Thus the distribution $P_{\rm ad}(\phi)$ will shift by the desired 
amount if and only if $\phi_{\rm het}$ is given by
\beq
\phi_{\rm het}(\hv,C) = \hv + g(C),
\eeq
for some arbitrary real function $g$ of $C$. Furthermore, it can be shown that 
for $H_{mn}$ to be real and positive we need $g(C^*)=-g(C)$.

\subsubsection{Adaptive Mark I Measurements}

The simplest choice is $g=0$. This corresponds to
\beq
\phi_{\rm I} = \hv = \arg A.
\eeq
That is, the phase estimate $\hv$ used in the feedback loop is also 
used as the final phase estimate. We call this the adaptive mark I 
measurement. In this case the POM is
\bqa
F^{\rm I}(\phi) &=& \int\!\!\! \int d^{2}C \,G^{\rm ad}(\phi,C). \\
&=& \int\!\!\! \int d^{2}C Q_{c}(C) \nn
\ket{\tilde{\psi}(e^{i\phi},e^{2i\phi}C)}
\bra{\tilde{\psi}(e^{i\phi},e^{2i\phi}C)}. 
\eqa

This POM can be easily evaluated in the number state basis using the 
definition (\ref{deftildepsi}). The result is in the form of 
Eq.~(\ref{genPOVM}) with the matrix $H$ given by
\bqa \label{HI}
H_{mn}^{\rm I} &=& \sum_{p=0}^{\lfloor m/2\rfloor} 
\sum_{q=0}^{\lfloor n/2\rfloor} \gamma_{mp}\gamma_{nq}  
\ip{ C^p (C^*)^q }_Q, \\
&=& \sum_{p=0}^{\lfloor m/2\rfloor} \sum_{q=0}^{\lfloor n/2\rfloor}
\gamma_{mp}\gamma_{nq} M^{p,q}.
\eqa
Here $\lfloor m/2\rfloor$ is the integer part of $m/2$ and
\beq
\gamma_{mp} = \frac{\sqrt{m!}}{2^p(m-2p)! p!}.
\eeq
This is an exact expression since the moments $M^{p,q}$ can be 
calculated exactly. 
It is not obvious from this definition $H^{\rm I}_{nn} = 1$ for all
$n$, but this can be verified computationally. 

The matrix $H^{\rm I}_{nn}$ is plotted in Fig.~2(c). It appears not 
greatly different from that for the heterodyne measurement. One 
 difference is that $H^{\rm I}_{1,m}=H^{\rm I}_{0,m}$ for all $m$, 
 and in particular that for $n,m \leq 1$, $H^{\rm 
I}_{n,m} = 1$. This is identical to the canonical measurement and as 
good as possible, as first revealed in Ref.~\cite{Wis95c}. This 
result shows that for very weak fields the adaptive mark I measurement 
is significantly better than the standard heterodyne technique. For 
moderate fields it is not significantly better (as Fig.~2 shows). As 
we will show later, for large fields it is very much worse. Evidently 
the adaptive mark I scheme is not the scheme we would choose for most 
practical situations in which the photon number per pulse is very large.

\subsubsection{Adaptive Mark II Measurements}

A generally better result can be obtained by considering a 
final phase measurements $\phi_{\rm ad}=\hv + g(C)$ with $g(C)\neq 0$. 
Recall the result \erf{weird} 
obtained above, that the probability of obtaining a measurement 
result is proportional to the squared inner product of the system 
state with a squeezed state 
\beq 
P(\alpha,\epsilon) \propto |\ip{\alpha,\epsilon | \psi}|^{2} .
\eeq
Here $\alpha,\epsilon$ are defined in terms of $A,B$ by 
Eqs.~(\ref{ss1}), (\ref{ss2}). We are interested in the case when the 
state $\ket{\psi}$ has a well-defined (but unknown) phase. Since any 
physical state will have a finite mean photon number this means that
must have a large coherent amplitude. As argued in Sec.~II~C, it is 
most likely that this coherent amplitude will be close to $\alpha$. 
Now in terms of the variables $\hv,C$ we have
\beq
\alpha = \frac{e^{i\hv}(1+C)}{1-|C|^{2}}
\eeq
This suggests the mark II phase estimate
\beq
\phi_{\rm II} = \arg\alpha = \hv + \arg(1+C).
\eeq
That is, we choose the function $g(C)$ so that
\beq
e^{ig(C)} = \sqrt\frac{1+C}{1+C^{*}}.
\eeq

With this choice 
\beq
F^{\rm II}(\phi) = \int\!\!\! \int d^{2}C \,G^{\rm ad}(\phi-\arg(1+C),C).
\eeq
The $H$ matrix is therefore
\bqa 
H_{mn}^{\rm II} &=& \sum_{p=0}^{\lfloor m/2\rfloor} 
\sum_{q=0}^{\lfloor n/2\rfloor} \gamma_{mp}\gamma_{nq} \times \nn \\
&& \ip{ \left(\frac{1+C}{1+C^{*}}\right)^{(n-m)/2} C^p (C^*)^q }_Q
\label{HII}
\eqa
Unfortunately $[(1+C)/(1+C^{*})]^{(n-m)/2}$ is not a polynomial in 
$C$ and $C^{*}$ so we cannot obtain an exact answer in terms of the 
known moments $M^{p,q}$. However from the definition (\ref{intC}) it 
is apparent that the modulus of the random variable $C$ is strictly bounded 
by unity. In fact $\ip{C}_{Q} = \ip{C^{*}}_{Q} = 
\ip{C^{*}C}_{Q}=1/3$, and all higher moments are smaller. Hence the 
MacLaurin series for $[(1+C)/(1+C^{*})]^{(n-m)/2}$ will converge rapidly 
and so can be well-approximated by a polynomial. Using an expansion to 
100 terms, we have evaluated this POM matrix elements for $n,m$ up to 
100. 

The matrix $H_{mn}^{\rm II}$ for $n,m$ up to 8 is shown in Fig.~2(d). From 
this it is apparent that the adaptive mark II scheme is generally much closer to 
a canonical measurement in this range than are either the heterodyne or 
adaptive mark I scheme. Indeed, all the matrix elements are above 
0.7, and all are greater than or equal to the heterodyne matrix 
elements. The only place where the adaptive mark II scheme is 
inferior to the adaptive mark I scheme is for very low photon 
numbers; $H_{01}^{\rm II} < 1$ unlike $H_{01}^{\rm I}$. We will show 
in the next section that the superiority of the mark II scheme over 
the other two schemes continues for large photon numbers, as 
quantified by the measured phase variance of various states.

\section{Phase Variance}

\subsection{Phase Variance and $H_{mn}$}

Because phase is a cyclic variable, the definitions of mean and 
variance which apply to the real line are not applicable. The 
sensible starting point for these two statistics for a cyclic variable 
with distribution $P(\phi)$ is
\beq
\mu = \int e^{i\phi}P(\phi) d\phi.
\eeq
The mean phase can then be defined to be
\beq
\bar{\phi} = \arg \mu,
\eeq
and the phase variance
\beq
V = |\mu|^{-2} - 1.
\eeq
It can easily be verified that these definitions go over to the usual 
ones appropriate for the real line when $P(\phi)$ is suitably 
localized (so that $1-|\mu|\ll 1)$. There are of course other 
definitions of the variance in terms  
of $|\mu|$ which would also give the correct limit 
\cite{BanPau69,Opa94}. The advantage of 
the one presented here is that it can be used to derive an 
uncertainty relation
\beq
4V \geq \left( \ip{a\dg a a\dg a}-\ip{a\dg a}\ip{a\dg a}\right)^{-1} ,
\eeq
as shown by Holevo \cite{Hol84}. This inequality holds for the 
variance of any 
$P(\phi)$ arising from a phase measurement conforming to the 
definition in Sec.~II~B.

Without loss of generality we can consider a system state 
\beq
\ket{\psi} = \sum_{n=0}^{\infty} \psi_{n} \ket{n},
\eeq
with real number state amplitudes $\psi_{n}$ so that it is guaranteed 
to have a mean phase of zero. The probability distribution from a 
phase measurement described by a POM (\ref{genPOVM}) with matrix $H$ is
\beq
P(\phi) =   \frac{1}{2\pi} \sum_{n,m=0}^{\infty} \psi_{m}\psi_{n} 
e^{i\phi(m-n)} H_{mn}.
\eeq
For such a system we have
\bqa
\mu &=& \sum_{n,m=0}^{\infty} \frac{1}{2\pi} \int d\phi e^{i\phi(m+1-n)}
\psi_{m}\psi_{n}  H_{mn} \\
&=& \sum_{n=0}^{\infty} \psi_{n+1}\psi_{n} H_{n+1,n} \label{mufrompsi}
\eqa
Thus the only part of $H$ which contributes to the 
phase variance is  the subdiagonal
\beq
H_{n+1,n} \equiv 1 - h(n).
\eeq

Although $h^{\rm II}(n)$ is not known exactly it was calculated to 
a very good approximation for $n$ up to 100, as explained above.
For heterodyne detection and adaptive mark I detection we 
have exact results and for a canonical phase measurement of 
course $h^{\rm can}(n)=0$. For large photon numbers it is more useful 
to have approximate asymptotic expressions for $h(n)$ for the three 
physically realizable schemes. These can be derived using 
semiclassical dyne detection theory \cite{WisKil97}. The results are
\bqa 
h^{\rm het}(m) &\simeq& (8m)^{-1} + O(m^{-2}), \label{h1}\\
h^{\rm I}(m) &\simeq& (8m^{1/2})^{-1} + O(m^{-1}) ,\label{h2}\\
h^{\rm II}(m) &\simeq&(16 m^{3/2})^{-1} + O(m^{-2}). \label{h3}
\eqa
As will be shown Sec.~IVB and IVC this leads to a clear superiority 
of the adaptive mark II scheme over 
the heterodyne scheme, and of the latter over the mark I 
scheme, for measuring the phase of states with large photon numbers.
Furthermore, it is shown at the end of App.~B that the adaptive mark 
II scheme is the best scheme for measuring large fields given the 
feedback algorithm (\ref{fa2}).

\subsection{Coherent States}

\label{seccs}

\subsubsection{Canonical}

A coherent state of mean phase equal to zero has coefficients
\beq
\psi_{n} = \exp(-\beta^{2}/2) \frac{\beta^{n}}{\sqrt{n!}}.
\eeq
Thus for a canonical measurement we can use \erf{mufrompsi} with 
$H_{mn}=1$ to get
\beq\label{canmucoh}
\mu = \exp(-\beta^{2}) \sum_{n=0}^{\infty} \frac{\sqrt{n} 
\beta^{2n}}{\beta n!}
\eeq
By expanding $\sqrt{n}$ in a Taylor series about $n=\beta^{2}$ while 
recognizing the moments of a Poisson distribution we obtain
\beq
\mu = 1 - \frac{1}{8\beta^{2}} - \frac{7}{128\beta^{4}} + 
O(\beta^{-6}).
\eeq
Thus the variance from a canonical measurement of the phase of a 
coherent state is
\beq \label{vccas}
V_{\rm coh}^{\rm can} = \frac{1}{4\beta^{2}} + \frac{5}{32\beta^{4}}
+ O(\beta^{-6}).
\eeq
This can be regarded as the intrinsic phase variance of a coherent 
state. In Fig.~3 we have plotted the exact result obtained 
numerically from \erf{canmucoh}, and the asymptotic result \erf{vccas} 
for $\beta$ from $1$ to $5$. The latter corresponds to a mean 
photon number of $25$, which is evidently large enough for the 
asymptotic results to hold quite well.

\subsubsection{Heterodyne}

For heterodyne detection we can use the exact expression \erf{heterH} 
to get
\beq
\mu = \beta \exp(-\beta^{2}) \sum_{n=0}^{\infty}
\frac{\Gamma(n+\frac32) \beta^{2n}}{\Gamma(n+2)\Gamma(n+1)}.
\eeq
In terms of confluent hypergeometric functions, this is
\beq
\mu = \beta \exp(-\beta^2) \frac{\Gamma(\frac{3}{2})}{\Gamma(2)}
        {}_{1}F_{1}(\smallfrac{3}{2};2;\beta^2). \label{exacthetVphi}
\eeq
Using the analogue to Euler's formula, 4.2(1) of \cite{Luk69}
asymptotic expansion
\beq
\mu = 1- \frac{1}{4\beta^2} - \frac{3}{32\beta^4} +O(\beta^{-6}).
\eeq
Thus the phase variance from a heterodyne measurement is
\beq \label{ashetcc}
V_{\rm coh}^{\rm het} 
= \frac{1}{2\beta^{2}} + \frac{3}{8\beta^{4}} +O(\beta^{-6}).
\eeq

To first (and almost to second) order this is twice that the canonical 
phase variance. The reason for this is apparent from the expression 
\erf{hetPOVM3} for the heterodyne POM. The probability distribution 
for a heterodyne phase measurement is
\bqa
P_{\rm het}^{\rm coh}(\phi) &=& \int_{0}^{\infty} |A|\, d(|A|) \, 
\bra{\beta} F^{\rm 
het}(|A|e^{i\phi}) \ket{\beta} \\
&=& \frac{1}{\pi} \int_{0}^{\infty} r\, dr |\ip{\beta|r 
e^{i\phi}}|^{2}.
\eqa
For $\phi$ close to the mean value of $0$ the integrand will be 
strongly peaked at $r \simeq \beta \gg 1$. Thus 
\beq \label{convol}
P_{\rm coh}^{\rm het}(\phi) \propto  |\ip{\beta|\beta 
e^{i\phi}}|^{2}.
\eeq
In other words, this distribution is approximately the convolution of 
the intrinsic phase distributions of two coherent states of 
amplitude $\beta$. Thus we expect the distribution to be 
approximately Gaussian, with a variance  double that of 
a canonical measurement. The exact result from \erf{exacthetVphi} and 
the asymptotic result \erf{ashetcc} are plotted on Fig.~3. The excess 
phase noise in the heterodyne result is because the measurement is not 
as good as the canonical result. In fact, we have
\beq \label{redhet}
V_{\rm coh}^{\rm het} - V_{\rm coh}^{\rm can} 
\simeq \frac{1}{4\beta^{2}} \simeq 2h^{\rm het}(\beta^{2}),
\eeq
where $h(m)$ is the asymptotic expression for 
$H_{m,m+1}-1$ given in \erf{h1}. The 
quantity in \erf{redhet}, which we will call the excess phase 
variance, is plotted in Fig.~4. From \erf{mufrompsi} it follows that, 
for states with a well-defined coherent amplitude,
the excess phase variance for any scheme is approximately
$2h(\beta^2)$.

\subsubsection{Mark I Adaptive}

It was shown in Ref.~\cite{WisKil97} that for a 
coherent state of amplitude $\beta \gg 1$ the adaptive mark I phase
$\hv$ can be approximated by a Gaussian random variable of mean zero 
and variance 
\beq
V_{\rm coh}^{\rm I} = \frac{1}{4\beta} + O(\beta^{-2}).
\eeq
This is plotted in Fig.~3 along with the exact result calculated 
from Eqs.~(\ref{mufrompsi}) and (\ref{HI}) truncated at $n=100$. 
This result shows that the adaptive mark I is far worse than a heterodyne 
measurement for large $\beta$. Indeed, to the order calculated, the 
phase variance is entirely due to the excess phase variance 
\beq \label{VcI}
V_{\rm coh}^{\rm I} - V_{\rm coh}^{\rm can} 
= \frac{1}{4\beta} + O(\beta^{-2}),
\eeq
This was the result used to obtain
\beq
h_{\rm I}(\beta^{2}) = \frac{1}{2}[V_{\rm coh}^{\rm I}- V_{\rm 
coh}^{\rm can} ]
= \frac{1}{8\beta} + O(\beta^{-2}),
\eeq
as recorded above in \erf{h2}. The asymptotic result (\ref{VcI}) and 
its exact value are plotted in Fig.~4. This shows that for small 
coherent states, with amplitude less than about 2, the mark I 
measurement introduces less excess noise than the heterodyne 
measurement. For $\beta = 5$ the asymptotic result is already a very good 
approximation.

\subsubsection{Adaptive Mark II}

For our final scheme we again used semiclassical techniques in 
Ref.~\cite{WisKil97} to show that $P^{\rm II}_{\rm coh}(\phi)$ was 
approximately Gaussian with a variance 
\beq
V_{\rm coh}^{\rm II} = \frac{1}{4\beta^{2}} + 
\frac{1}{8\beta^{3}} + O(\beta^{-4}).
\eeq
Like the canonical result, this is dominated by the intrinsic phase 
noise of the coherent state.
This asymptotic result, and the exact result from 
Eqs.~(\ref{mufrompsi}) and (\ref{HII}), are plotted in Fig.~3. 
The excess phase noise in this case is
\beq
2h_{\rm II}(\beta^{2}) = 
V_{\rm coh}^{\rm II} - V_{\rm coh}^{\rm can} 
= \frac{1}{8\beta^{3}} + O(\beta^{-4}),
\eeq
which is far below that of the other two dyne schemes. This 
asymptotic result, and the exact excess phase variance, are plotted in 
Fig.~4. Once again, the asymptotic behaviour is evident for $\beta = 
5$.

\subsection{Phase-optimized states}

From the coherent state results, the marked superiority of the 
adaptive mark II measurement over the standard techniques is apparent 
only from considering the excess phase variance. A more direct 
measure is the {\em minimum} phase variance for each measurement 
scheme. In this measure, the state is optimized for each scheme, and 
is subject to the constraint of having a maximum photon number $N$. 
That is to say we have to optimize the unit-norm real vector 
$(\psi_{0},\psi_{1},\ldots \psi_{N})$ 
so as to {\em maximize}
\beq
\mu = \sum_{n=0}^{N} \psi_{n+1}\psi_{n} [1-h(n)] .
\eeq
This can be rewritten as
\beq
\mu = \frac12 \sum_{m,n=0}^{N} \psi_{m} J_{mn} \psi_{n}
\eeq
where
\beq
J_{mn} = \half [1-h(n)] \delta_{m,n+1} + 
\half [1-h(m)] \delta_{m,n-1}.
\eeq
The problem of maximizing  $\mu$ thus reduces to 
that of finding the largest eigenvalue $\lambda_{\rm max}$ of
the real symmetric matrix $J$. 
Since we have $h(n)$ for all schemes up to $n=100$ this can be done 
for a maximum photon number $N$ up to $100$. 

For the canonical case with $h(m)=0$ 
the eigenvalue can be found exactly to be
\beq
\lambda_{\rm max} = \cos\left(\frac{\pi}{N+2}\right)
\eeq
so that
\beq
V_{\rm min}^{\rm can} = \tan^{2}\left(\frac{\pi}{N+2}\right) 
= \frac{\pi^2}{N^2} -4\frac{\pi^2}{N^3}+O(N^{-4}).
\eeq
For the dyne measurements there is no analytical solution but a 
numerical solution is easily obtained. The results are plotted in 
Fig.~5. This clearly shows the same order as established for coherent 
states with large photon numbers: the adaptive mark II measurement is 
best, followed by heterodyne, followed by adaptive mark I.

Also plotted in Fig.~5 are the asymptotic results for the three dyne 
measurements. These were obtained in Ref.~\cite{WisKil97} using the 
asymptotic results for $h(n)$ of Eqs.~(\ref{h1})--(\ref{h3}). The 
results are most easily expressed by noting that these functions 
$h(n)$ can all be written as
\beq
h^{\rm dyne}(n) = c n^{-p}
\eeq
for some positive power $p \geq 1/2$ and positive coefficient $c$ of order 
unity. From this we got
\beq \label{pl}
V_{\rm min}^{\rm dyne} \approx
2cN^{-p} + (-z_{1})(2cp)^{2/3}N^{-2(1+p)/3},
\eeq
were $z_{1} \approx -2.338$ is the first zero of the Airy function.
The leading term here is simply equal to $2h(N)$. This is essentially 
the excess noise introduced by the measurement, just as 
$2h^{\rm dyne}(\beta^{2})$ was for the coherent state. In this case the 
intrinsic noise (the second term) varies between the different 
schemes because  the state is optimized for each measurement.

From Fig.~5 it is apparent that 
the exact numerical results are approaching this asymptotic 
result for the heterodyne and mark I measurements. However the mark II 
exact results are a long way from the asymptotic results even with 
$N=100$. This is actually not surprising. A simple calculation carried 
out in Ref.~\cite{WisKil97} suggested that the asymptotic results 
would only become valid for 
\beq
N \agt N_{\rm as} = \left( \frac{10^{3}}{2cp} \right)^{1/(2-p)}.
\eeq
For an adaptive mark I measurement we have $N_{\rm as} = 400$; for 
heterodyne $N_{\rm as} = 4000$; and for adaptive mark II $N_{\rm as} 
\approx 3 \times 10^{7}$. Evidently these requirements are 
overly-conservative (as mooted in our earlier paper). Nevertheless, 
it does explain why the minimum adaptive mark II phase variance 
 is a long way from reaching its asymptote for $N=100$. This 
 underlines the usefulness of the approximate asymptotic results. An 
 exact numerical solution with $N=10^{7}$ would be severely impractical. 
 It also points out the danger of trying to derive power laws such as 
 Eq.~(\ref{pl}) from numerical data for moderate photon numbers of a few 
 hundred, as done by D'Ariano and Paris in Ref.~\cite{DArPar94}. 
 A detailed comparison with their results for heterodyne detection for 
 optimized states with a fixed mean photon number will appear in a 
 future paper.

\section{Phase Probability Distributions}

\subsection{$P(\phi)$ for coherent states}

\label{secppcs}

Although the semiclassical theory of Ref.~\cite{WisKil97} has proven 
invaluable for calculating the asymptotic phase variance for states of 
large photon number, it cannot readily yield the total phase 
distribution $P(\phi)$. This is the quantity that is needed for 
a proper analysis of optical communication based on encoding 
information in the  phase of single-mode pulses. For a communication 
system there are certain phases which one would be expecting to 
receive, so what matters is not the mean-square error in the phase 
measurement, but the probability for mistaking one phase for another. 
This depends on the total $P(\phi)$, which requires knowledge of the 
full matrix $H_{mn}$:
\beq
P(\phi) =   \frac{1}{2\pi} \sum_{n,m=0}^{\infty} \rho_{mn} 
e^{i\phi(m-n)} H_{mn},
\eeq
where $\rho_{mn}$ is the density matrix for the system state in the 
photon number basis.

Before calculating  probabilities of error it 
is informative simply to plot 
$P(\phi)$ for the various schemes with the system in a coherent state. 
In Fig.~6 we plot $\log P_{\rm coh}(\phi)$ versus $\phi$ for various values of 
coherent amplitude $\beta$. 
One thing is clear: the 
canonical $P(\phi)$ is best by any definition. For small 
coherent amplitudes  
the adaptive mark I case is the best dyne measurement, 
and is almost indistinguishable from the canonical measurement. As 
$\beta$ becomes larger the peak of $P^{\rm het}_{\rm coh}(\phi)$ 
becomes sharper and taller than that of $P^{\rm I}_{\rm coh}(\phi)$. 
The peak of $P^{\rm II}_{\rm coh}(\phi)$ becomes sharper and taller 
still, and for moderate $\beta$ is indistinguishable 
and that of $P^{\rm can}_{\rm coh}(\phi)$. All of the curves are 
inverted parabolas for small $\phi$, indicating that the 
distributions $P(\phi)$ are approximately Gaussian. 

All of these features could be predicted from the above results. What is 
unexpected is the shape of the tails of the curves. First, as $\beta$ 
increases, $P^{\rm can}_{\rm coh}(\phi)$ ceases to fall monotonically 
with distance from $\phi=0$, but suddenly reverses 
 at $\phi \approx 1$ and has a broad local maximum at $\phi=\pi$. 
The heterodyne distribution has no such reversal, but 
nevertheless levels out and approaches the 
canonical value at $\phi = \pi$. The adaptive mark I case is also 
apparently smooth, but has much higher tails than the canonical 
heterodyne distributions. The big surprise is the adaptive mark II 
distribution. Like the canonical distribution it reverses (although 
smoothly) and has a broad local maximum at $\phi= \pi$. But the value of 
$P^{\rm II}_{\rm coh}(\pi)$ is actually the largest of all four 
schemes! In fact, for large $\beta$, $P^{\rm II}_{\rm coh}(\pi)$ 
closely follows $P^{\rm can}_{\rm coh}(\phi)$ 
 until it reaches a floor, which is roughly the same as that of 
 $P^{\rm I}_{\rm coh}(\phi)$. 
 
These features are not easy to explain from the matrix elements $H_{mn}$. 
For example, the ratio of the probability density at $\phi=\pi$ to 
that at $\phi= 0$ is given by
\beq
\frac{P(\pi)}{P(0)}= \frac{\sum_{mn} H_{mn} (-1)^{m-n} 
\beta^{m+n}/\sqrt{n!m!}}{\sum_{mn} H_{mn} \beta^{m+n}/\sqrt{n!m!}}.
\eeq
Evidently this ratio depends crucially on the relative values of the 
matrix elements $H_{mn}$ for $m,n \sim \beta^{2}$. In particular, 
 just because  $H^{a}_{mn} \geq H^{b}_{mn}\, \forall\, m,n$ it does not follow that
$P^{a}(\pi) \leq P^{b}(\pi)$.
That is, a measurement with a POM closer to the canonical POM, in 
the sense of having all elements of $H_{mn}$ closer to unity, does not 
guarantee an unambiguously better phase probability distribution.

\subsubsection{Heterodyne Measurements}

For heterodyne detection we can find an expression for 
 $P(\pi)$ analytically. Recall that in this case the POM is
\beq
G'_{\rm coh}(\alpha)d^{2}\alpha = \frac{1}{\pi} \ket\alpha \bra\alpha 
d^{2}\alpha,
\eeq
where $\ket\alpha$ is a coherent state 
and the phase estimate is $\phi = \arg\alpha$. Clearly then the 
probability to obtain $\phi = \pi$ is
\bqa
P^{\rm het}_{\rm coh}(\pi) &=& 
\frac{1}{\pi}\int_{0}^{\infty} r dr |\ip{\beta |{-r}} |^{2},\\
&=& \frac{1}{\pi}\int_{0}^{\infty} r dr \exp\bl - (\beta+r)^{2} \br .
\eqa 
This integral can be evaluated in terms of the error function, but 
for $\beta\gg 1$ it is well approximated by
\beq
P^{\rm het}_{\rm coh}(\pi) = \frac{1}{4 \pi \beta^{2}}\exp(-\beta^{2}).
\eeq
It can be verified from Fig.~6 that this is a very good approximation 
even for $\beta = 5$.
For very large $\beta$ the most important contribution is the 
$\exp(-\beta^{2})$ term. This scaling can be expressed as
\beq
\log P^{\rm het}_{\rm coh}(\pi)  \simeq -\beta^{2}.
\eeq

\subsubsection{Adaptive measurements}

For the adaptive measurements we can also determine $P(\pi)$ by 
returning to the POM 
\bqa 
G^{\rm ad}(\hv,C)\, d\hv\, d^{2}C &=&
\frac{d\hv}{2\pi} d^{2}C Q_{c}(C) \times \nn \\
&& \ket{\tilde{\psi}(e^{i\hv},e^{2i\hv}C)}
\bra{\tilde{\psi}(e^{i\hv},e^{2i\hv}C)},
\label{adPOVMagain}
\eqa
where 
\beq
\ket{\tilde{\psi}(e^{i\hv},e^{2i\hv}C)}
= \exp\bigl( \mbox{$\frac12$} e^{2i\hv}C {a^\dagger}^2  + e^{i\hv}
a^\dagger \bigr) \ket{0}.
\eeq
For a coherent state $\ket{\beta}$ with $\beta$ real 
the probability density is
\bqa
P^{\rm ad}_{\rm coh}(\hv,C) &=& \frac{Q_{c}(C)}{2\pi} 
|\bra{\beta}\tilde{\psi}(e^{i\hv},e^{2i\hv}C) \rangle |^{2}\label{forapB} \\
&=& \frac{Q_{c}(C)}{2\pi}  \exp\left(-\beta^{2} + {\rm Re}[e^{2i\hv}C\beta^{2}
+ 2e^{i\hv}\beta] \right). \nn
\eqa

Consider first the adaptive mark I scheme for which $\phi = \hv$. 
The ratio of $P_{\rm coh}^{\rm I}(\pi)$ to $P_{\rm coh}^{\rm I}(0)$ is
\bqa
\frac{P_{\rm coh}^{\rm I}(\pi)}{P_{\rm coh}^{\rm I}(0)} &=& 
\frac{\int\!\!\int d^2{C} P_{\rm coh}^{\rm ad}(\pi,C)}
{\int\!\!\int d^2{C} P_{\rm coh}^{\rm ad}(0,C)} \label{0pi}\\
&=& \frac{\int\!\!\int d^2{C} Q_{c}(C)
 \exp\left(-\beta^{2} + {\rm Re}[C\beta^{2}]
- 2\beta \right)}{\int\!\!\int d^2{C} Q_{c}(C)
 \exp\left(-\beta^{2} + {\rm Re}[C\beta^{2}]
+ 2\beta \right)} \nn \\
&=& \exp(-4\beta).
\eqa
Now since $P^{\rm I}_{\rm coh}(\phi)$ is approximately Gaussian we 
have $P^{\rm I}_{\rm coh}(0) = (2\pi V^{\rm I}_{\rm coh})^{-1/2} = 
(\pi/4\beta)^{-1/2}$, so that
\beq
P_{\rm coh}^{\rm I}(\pi) \simeq \sqrt\frac{4\beta}{\pi} \exp(-4\beta).
\eeq
This agrees excellently with the numerical result plotted in Fig.~6 
for $\beta = 5$. For very 
large $\beta$ the dominant term is obviously the exponential, 
which we can express by the equation
\beq
\log P_{\rm coh}^{\rm I}(\pi) \simeq -4\beta.
\eeq

For the adaptive mark II scheme we expect  the tail of the 
distribution to be at least as high as that for the adaptive mark I 
case, which is what is indeed seen. That is because
\beq
\phi = \hv + \arg(1+C),
\eeq
and $\arg(1+C)$ lies between $-\pi/2$ and $\pi/2$. Thus irrespective 
of $C$, a result $\hv\approx \pi$ 
in the tail of the distribution of the mark I 
measurement must also give a result $\phi$ in the tail of the mark II 
measurement. By this crude argument we would also expect the log of 
the tail of 
the distribution of the mark II measurement to scale in the same way:
\beq
\log P_{\rm coh}^{\rm II}(\pi) \simeq - 4\beta.
\eeq
Clearly the relative disparity between the height of tails of 
the adaptive measurements and those of the heterodyne or canonical 
measurements will continue to increase as $\beta$ increases. 
A discussion about the reason for this disparity is to be found in 
App.~B.

\subsection{$M$--ary encoding with coherent states}

As stated above, one reason for wishing to know the complete phase 
probability distributions, including the tails, is for calculating 
the effectiveness of the various schemes for digital communication 
using phase encoding. The canonical and heterodyne POMs have been
examined before by Hall and Fuss \cite{HalFus91}. Here we follow their approach, 
and consider  $M$--ary encoding; that is,
the transmission of data as the string of $M$--ary digits $\{0,1,...,M-1\}$.
Each digit is represented by a rotated version of some single quantum state
$\ket{\psi}$ whose phase distribution is peaked about zero.
The digit $n$ is encoded as $\exp({\frac{2in\pi}{M}a^\dagger
a})\ket{\psi}$.
The receiver makes a phase measurement (as defined in Sec.~IIB) 
on this state and infers from the
result which digit was sent. That is, a result
$\phi$ in the interval $2\pi n/M \pm \pi/M$  is interpreted as the digit
$n$.

 The essential measure of any mode of digital communication is the
probability that an error occurs. For each of the four measurement schemes
we have calculated the minimal probability of error that may be achieved for
each of two types of transmitted states. The first type is coherent 
states. These are important because, with the exception of squeezed
states \cite{WalMil94},
they are perhaps the only pure single-mode 
quantum states that can be produced readily enough
to be considered for communication applications.  

Under the decoding scheme described above the probability of error is
independent of the digit encoded. For the zero state it is
\beq \label{E}
E = \int_{\pi/M}^{2\pi-\pi/M} P(\phi) d\phi 
\eeq
It is easy to see that $E$ is the expectation value of the positive
operator $F_E = 1-F_C$ where 
\beq
\label{ErrorPOVM}
F_{C}=\sum_{n,m=0}^\infty \frac{\sin
[\pi(m-n)/M]}{\pi(m-n)}  H_{m,n} |m \rangle\langle n|.
\eeq
Using this operator, the
expansion of a coherent state in terms of number states, and the values of
$H_{m,n}$ for $0\leq m,n \leq 100$ computed earlier, one may easily 
determine the
probability of error for coherent states with small $\beta$.

We can find approximate asymptotic analytic expressions for $E$ by returning to 
\erf{E}. The logarithm of $E$ will be well approximated by the 
logarithm of the largest value of the integrand in \erf{E}. Since 
$P(\phi)$ for coherent states is approximately monotonically 
decreasing from $\phi=0$ to $\phi=\pi$ for all schemes, we can thus say
\beq
\log E_{\rm coh} \approx \log P_{\rm coh}(\pi/M).
\eeq
To proceed further we make the approximation that
 $P_{\rm coh}(\phi)$ is Gaussian until it hits the floor value $P(\pi)$.
That is,
\beq
\log P_{\rm coh}(\phi) \approx \max\{ -\phi^{2}/2V_{\rm coh}, \log P_{\rm 
coh}(\pi) \},
\eeq
so that
\beq
\log E_{\rm coh} \approx - \min \left\{ \frac{\pi^{2}}{2M^{2}V_{\rm coh}}, 
\log P_{\rm coh}(\pi) \right\}.
\eeq

From the results of Sec.~\ref{seccs} and Sec.~\ref{secppcs} we can 
evaluate this expression for the probability of error for the various 
schemes. 
\bqa
\log E_{\rm coh}^{\rm can} &\approx& - \beta^{2} \min \left\{ 2 
(\pi/M)^{2}, 1  \right\} \label{ap1}\\
\log E_{\rm coh}^{\rm het} &\approx& - \beta^{2} \min \left\{  
(\pi/M)^{2}, 1  \right\} \label{notsg}\\
\log E_{\rm coh}^{\rm I} &\approx& - \beta \min \left\{ 2 
(\pi/M)^{2}, 4  \right\} \label{ap2}\\
\log E_{\rm coh}^{\rm II} &\approx& - \beta \min \left\{ 2 
\beta (\pi/M)^{2}, 4  \right\}. \label{ap3}
\eqa
As long as $\beta > 2(M/\pi)^{2}$ we have the simple results that 
$-\log E$ scales quadratically with $\beta$ for canonical and 
heterodyne measurements, and linearly with $\beta$ for the two 
adaptive measurements. For $\beta < 2(M/\pi)^{2}$ the adaptive mark II 
measurement scales quadratically.

From Fig.~6 it is evident that the approximation of $P(\phi)$ as a 
Gaussian plus a constant tail is poorest for the heterodyne 
measurement. Thus we would not expect the expression (\ref{notsg}) to 
be particularly good. However for this measurement 
scheme we can find the following expression for $E$:
\beq
1-E^{\rm het}_{\rm coh} = \frac{1}{\pi}\int_0^\infty \int_0^{ay}
      e^{-(\beta-x)^2-y^2} dx dy
\eeq
where $a=\cot(\pi/M)$. After quite some effort this yields the 
asymptotic expression
\bqa
\log(E^{\rm het}_{\rm coh}) &\simeq& -\beta^2 / (1+a^2) 
     + \log\left( \frac{ (1+a^2)^5-a^{10} }{ \sqrt{\pi} (1+a^2)^{9/2}} 
     \right)  \nn \\
     && + \,\log(\beta)  + O(\beta^{-1}). \label{muchbet}
\eqa
The leading term of this differs from the above result (\ref{notsg}) by 
at most $25\%$ (for $M=3$) and approaches it for large $M$.
The full expression (\ref{muchbet}), and the above approximate expressions 
(\ref{ap1}),(\ref{ap2}) and (\ref{ap3}) are plotted as a function 
of $\beta$ in Fig.~7 for $M=4$. Also plotted are the exact numerical 
calculations of the probability of error. The expression 
(\ref{muchbet}) is evidently a very good approximation. The other 
analytical expressions match quite well the slopes of the curves, but 
are displaced vertically. For large $\beta$ the slope is of course 
the more important feature, and it is interesting that \erf{ap3} does 
correctly predict the change from quadratic to linear behaviour  
of $\log E_{\rm coh}^{\rm II}$ at $\beta \approx 2(4/\pi)^{2} \approx 
3.24$. 

From the asymptotic results it is clear that
 for large $\beta$ the adaptive mark II 
measurement has a higher probability of error than heterodyne 
detection. Specifically, for $M > 3$ the cross-over point is at 
\beq 
\beta \approx 4(M/\pi)^{2}.
\eeq
For $M=4$ this is $\beta \approx 6.48$, which agrees well with the 
numerical data in Fig.~7. At this point the error is
\beq
\log E_{\rm coh} \approx -16(M/\pi)^{2}.
\eeq
Thus depending on whether the acceptable error level is less than or 
greater than this amount, the best dyne measurement scheme to use (in the 
sense of requiring the least energy $\hbar\omega \beta^{2}$ per pulse)
will be heterodyne or adaptive mark II respectively.

\subsection{$M$--ary encoding with optimal states}

In this section we consider the probability of error for optimized 
states subject to a maximum-photon-number constraint.
Since the probability of error is
\beq
E = \bra{\psi}1-F_{C}\ket{\psi},
\eeq
it is readily seen  that the problem of finding the
minimal probability of error for states of the form
$
\sum_{n=0}^N c_n |n\rangle
$
is precisely that of finding the largest eigenvalue of the matrix 
formed by truncating the number-state matrix for $F_{C}$ of \erf{ErrorPOVM}. 
For small $N$ this eigenvalue
problem can be solved using MATLAB and the  
$H_{mn}$ matrices computed earlier.
 
Figure 8 depicts the results for quaternary ($M=4$) encoding.
It is clear from this graph that the log of the $E_{\rm opt}$
for optimized states 
has the same sort of dependence of the maximum photon number $N$ as 
the log of $E_{\rm coh}$ has on the mean photon number 
$\beta^{2}$. That is, for large $N$, the heterodyne and canonical 
measurements scale linearly with $N$ (with the latter having the 
greater slope) while the adaptive measurements scale as the square 
root of $N$ (with the adaptive mark II having the greater slope). Once 
again the adaptive mark II measurement is the best realizable 
measurement for moderate $N$, while the heterodyne measurement 
becomes superior for large $N$. We would expect the cross-over point 
to scale as $M^{4}$, and for $M=4$ the numerical data shows that it is
at $N \approx 64 \approx 25 (M/\pi)^{4}$.

\section{Discussion}

In this paper we have presented the exact quantum theory of two 
adaptive phase measurements. From this we have confirmed the 
semiclassical results obtained in Ref.~\cite{WisKil97}.  In 
particular, the phase variance from our adaptive mark II 
phase measurement is always less than that from a standard phase measurement 
(such as heterodyne detection). We have also applied our theory to an 
area inaccessible to the semiclassical theory, that is the complete 
shape of the probability distribution for the measured results $\phi$. 
We find that the adaptive  measurement phase 
probability distributions have surprisingly high tails. 
This has the consequence that the adaptive measurement is not 
necessarily better than standard phase measurements when it comes to 
communication using $M$--ary encoding of data in the phase of states.

The fact that the adaptive phase measurement is not necessarily superior 
to the standard phase measurement for $M$--ary phase encoding does not 
mean that it is a poor phase measurement, or that adaptive 
measurements in general are not useful. After all the situation of 
$M$--ary encoding does  not really 
call for a phase measurement; rather it calls for a measurement 
which can distinguish as well as possible 
between a finite number of known different (but 
not orthogonal) states. For the case of binary 
phase encoding using coherent states (with phases $0$ and $\pi$),
there is an adaptive measurement which has been 
known for some time \cite{Dol73} which 
distinguishes these possible states as well as quantum mechanics 
allows. It is only when $M \sim N$, where $N$ is the 
mean photon number of the states, that the measurement required is 
really a phase measurement. In this limit the variance of the 
distribution is the important factor, and the adaptive mark II phase 
measurement always gives a lower error rate than standard detection.

Although the asymptotics for the  
phase variance of the adaptive schemes were 
already known from the semiclassical theory of Ref.~\cite{WisKil97}
the quantum theory presented here sheds knew light on these results 
and allows us to probe new issues. For example, what is the ultimate 
limit on the phase noise introduced by an adaptive phase measurement? 
In other words, how closely is it possible to approximate a canonical 
phase measurement by using a measurement involving dyne measurements 
(that is measurements using photodetection 
and a local oscillator with arbitrary time-varying phase)? 
Although we cannot answer this question at this stage, we can show 
that there is a lower bound on the amount of excess noise. This lower 
bound is not 
due to imperfections such as a finite local oscillator or inefficient 
detectors, but is a fundamental limitation of the method of measurement 
via photodetection. We proceed by using the analysis in App.~B. 

It was shown in App.~B that the probability for obtaining a 
particular phase $\phi$ is determined largely by the maximum overlap 
between the system state and any of the 
pure states which contribute to the probability operator $F(\phi)$ for 
that phase. For dyne measurements, these pure states are squeezed 
states. As a result of this, the variance of the measured phase probability 
distribution will be (to a good approximation) equal to 
the true (canonical) phase variance of the system 
plus the phase variance of the maximum-overlap pure state. 
Furthermore, it was shown in App.~B that in order to obtain a large 
overlap, the maximum-overlap squeezed state 
must have a well-defined coherent amplitude 
roughly equal to the coherent amplitude of the system. 

From these considerations we can conclude that if the 
system has roughly $N$ photons, then the excess phase variance will be 
approximately that of a squeezed state with a mean photon number of $N$. 
Now the minimum (canonical) phase variance of a squeezed state with a mean photon 
number of $N$ has been investigated by Collett \cite{Col93}, who found the 
asymptotic result
\beq \label{lb2}
V^{\rm can}_{\rm ss} \geq \frac{\log N}{4 N^{2}}.
\eeq
This represents a lower bound on the excess phase variance introduced by 
any dyne measurement. So, for example, if $N$ is sufficiently large 
then the minimum measured phase variance for a state with at most $N$ 
photons would be
\beq \label{minmin}
V^{\rm dyne}_{\rm min} \geq \frac{\log N}{4 N^{2}}.
\eeq

This lower bound should is a long way below the variance achieved by 
the adaptive mark II scheme presented here, for which
\beq 
V^{\rm II}_{\rm min} \simeq \frac{1}{8N^{3/2}},
\eeq
which itself is a long way below the the variance achieved by standard 
measurements, namely
\beq
V^{\rm het}_{\rm min} \simeq \frac{1}{4N}.
\eeq
In fact, the lower bound (\ref{minmin}) is 
very close to the absolute lower limit set by canonical 
measurement \cite{fn2}
\beq
V^{\rm can}_{\rm min} \simeq \frac{\pi^{2}}{N^{2}}.
\eeq
Exactly how close one can come to the lower bound (\ref{lb2}) by using 
a different feedback algorithm is a 
matter for future research.

\acknowledgements

HMW  would like to thank  the Australian Research Council, and 
RBK the W.H.~Pickering Fellowship,   for financial support.

\appendix

\section{The ostensible moments of $C$}

Following the text, we denote the ostensible moments of $C$ as
\beq
M^{n,m}_{v} = \langle C_{v}^{n} {C_{v}^{*}}^{m} \rangle_{Q}.
\eeq
Using the rules of \ito calculus to evaluate 
\beq
d M^{n,m}_{v} = \ip{ (C_{v}+dC_{v})^{n} (C_{v}^{*}+dC_{v}^{*})^{m} 
- C_{v}^{n}{C_{v}^{*}}^{m}} 
\eeq
we find from \erf{deC} 
\beq
\frac{dM^{n,m}_{v}}{dv} =
 -\frac{2(n-m)^{2}}{v} M_{v}^{n,m} + n M_{v}^{n-1,m} + m M_{v}^{n,m-1}
\eeq
Since $M^{0,0}_v \equiv 1$ these equations may be solved recursively to find,
\beq
M^{n,m} = \frac{n M^{n-1,m} + m M^{n,m-1}}{2(n-m)^{2} + n + m}.
\eeq
Recall that by convention $M^{n,m} = M^{n,m}_{1}$. For $n$ or $m$
For $n$ or $m$ 
equal to zero this recurrence relation can be solved to get
\beq
M^{n,0} = M^{0,n} = \frac{1}{(2n+1)(2n-1)\ldots 1} = 
\frac{1}{(2n+1)!!}.
\eeq
These boundary values allow us to rapidly compute all the desired 
moments $M^{n,m}$.

\section{The tails of the distributions}

The 
reason for the different scaling of the tails of the adaptive 
measurements compared to the heterodyne measurement can be 
understood as follows. For heterodyne detection the dominant term 
is the inner product of the system 
state $\beta$ with the coherent state $\ket{-r}$ for  $r=0^{-}$. This
maximizes the overlap while still maintaining $\phi = \arg r = \pi$:
\beq
\log P_{\rm coh}^{\rm het}(0) \simeq \log |\ip{\beta|0}|^{2} 
= -\beta^{2}.
\eeq
For the adaptive mark I technique the overlap will be with a squeezed state 
$\ket{\alpha,\epsilon}$, where (using $\phi = \hv = \pi$)
\bqa
\alpha &=& - \frac{1+C}{1-|C|^{2}} \\
\epsilon &=& - \frac{C {\rm atanh}|C|}{|C|} .
\eqa
The problem is to determine the value of $C$ which maximizes this 
overlap.

It is not difficult to see that the value of $C$ we seek will be real 
and positive. In this case
\bqa
\alpha &=& -(1-C)^{-1} \\
\epsilon &=& -{\rm atanh}C
\eqa
This describes a squeezed state centred at $x=-2/(1-C)$ with an 
$x$-variance
\beq
 \exp(-2\epsilon) = \frac{1+C}{1-C} .
 \eeq
The overlap between $\ket{\beta}$ and $\ket{\alpha,\epsilon}$ is
\bqa
|\ip{\beta|\alpha,\epsilon}|^{2} &=& 
\frac{\exp\left[ - (1+\tanh \epsilon)(\beta+\alpha)^{2}\right]}{\cosh 
\epsilon}\\
&\simeq& 
\frac{\exp\left[-(1-C)\left(\beta + 
\frac{1}{1-C}\right)^{2}\right]}{\sqrt{1-C^{2}}} 
\eqa
Ignoring the negligible $\sqrt{1-C^{2}}$, this expression is 
maximized for
\beq
1-C = \beta^{-1}.
\eeq 
This implies $\alpha = - \beta$ and $\exp(-2\bar\epsilon) \simeq 
2\beta$. Substituting this in gives
\beq
\log P_{\rm coh}^{\rm I}(\pi) \simeq \log |\ip{\beta|\alpha,\epsilon}|^{2}
\simeq -4\beta,
\eeq
as obtained in the body of the paper.

This derivation in the appendix shows that the reason for the high 
tails of the adaptive distributions is the large 
 $x$-variance of the squeezed state $\ket{\alpha,\epsilon}$, 
giving it a much larger overlap with $\ket\beta$ 
than has $\ket{0}$ (from the heterodyne measurement). 
Although this large squeezing is responsible for the high tails, it 
is also what allows the narrow peak of the adaptive mark II measurement. 
This can be seen as follows. 

The most likely result for the adaptive mark II case is $\phi=\hv + 
\arg(1+C) = 0$. This is obviously most likely to occur for $\hv = 0$, 
in which case the only difference is that
\beq
\alpha =  \frac{1+C}{1-|C|^{2}} 
\eeq
One again it is easy to see that the maximum overlap will be for 
$C\approx 1$. The overlap in this case is
\beq
\log |\ip{\beta|\alpha,\epsilon}|^{2} 
\simeq -(1-C)\left(\beta - 
\frac{1}{1-C}\right)^{2}
\eeq
This is maximized (with a value of zero) at exactly the same $C = 
1-\beta^{-1}$. This gives $\alpha = \beta$ as expected, and the same 
$x$-variance. 

In this case what is of more interest is the 
 $y$-variance
\beq
\exp({2\epsilon}) \simeq (2\beta)^{-1}.
\eeq
The {\em intrinsic} phase variance of this squeezed state is thus
\beq
V_{\rm ss} \simeq \frac{\ip{y^{2}}}{\ip{x}^{2}} = 
\frac{\exp({2\bar\epsilon})}{(2\beta)^{2}} \simeq 
\frac{1}{8\beta^{3}}.
\eeq
This is precisely equal to the asymptotic expression for the excess variance 
\beq
V^{\rm II}_{\rm coh} - V^{\rm can}_{\rm coh} \simeq \frac{1}{8\beta^{3}}.
\eeq
The reason for this is that the measured phase distribution is at least as wide as 
a convolution of 
the true (canonical) phase distribution of the state with the true 
phase distribution of the most likely POM. This is completely 
analogous to the argument centred around \erf{convol} for the heterodyne case. 
For the adaptive mark I measurement the measured distribution is 
actually much wider, but the above calculation 
shows that for the adaptive mark II measurement 
all of the introduced noise is due to the quantum uncertainty in the 
states making up the POM. Thus the mark II phase estimate is, for large 
fields, the best possible estimate given the feedback algorithm (\ref{fa2}).

\begin{figure}
	\caption{\narrowtext Diagram for the experimental apparatus for making an 
	adaptive phase measurement. Thin dashed lines indicate light rays and 
	the thin continuous line labeled BS represents a 50/50 beam splitter. 
	Medium lines represent electro-optic devices: 
	photodetectors (PD) and an electro-optic phase modulator (EOM). Thick 
	lines represent electrical components: a subtractor, a multiplier, 
	an integrator, a signal generator (SG), a signal processor, and a 
	digital read out giving the measured value of $\phi \in [0,2\pi)$.
	The necessity for these 
	particular electrical elements alone is a consequence of the 
	feedback algorithm explained in Sec.~III~B.
	}
	\protect\label{fig1}
\end{figure}

\begin{figure}
	\caption{\narrowtext Plot of the $H$ matrix which defines the POM for phase 
	measurements as in Eq.~\ref{genPOVM}, for the four schemes (a) 
	canonical, (b) heterodyne, (c) adaptive mark I, and (d) adaptive mark 
	II.
	}
	\protect\label{fig2}
\end{figure}
 
\begin{figure}
	\caption{\narrowtext Plot of the exact (points) and asymptotic (lines) 
	expressions for the phase 
	variance $V_{\rm coh}$ 
	of a coherent state of amplitude $\beta$ versus $\beta$ 
	under the four schemes:  
	canonical ($*$ and solid line); heterodyne ($\circ$ and 
	dotted line); adaptive mark I (+ and dash-dot line); and 
	adaptive mark II ($\times$ and dashed line).
	}
	\protect\label{fig3}
\end{figure}
 
\begin{figure}
	\caption{\narrowtext Plot of the exact (points) and asymptotic (lines) 
	expressions for the excess phase 
	variance $V_{\rm coh}- V_{\rm coh}^{\rm can}$ 
	of a coherent state of amplitude $\beta$ versus $\beta$ 
	under the three dyne schemes:  heterodyne ($\circ$ and 
	dotted line); adaptive mark I (+ and dash-dot line); and 
	adaptive mark II ($\times$ and dashed line).
	}
	\protect\label{fig4}
\end{figure}
 
\begin{figure}
	\caption{\narrowtext Plot of the exact (points) and asymptotic (lines) 
	expressions for the minimum phase 
	variance $V_{\rm min}$ 
	of the optimal state with at most $N$ photons versus $N+1$ 
	under the four schemes:  
	canonical ($*$ and solid line); heterodyne ($\circ$ and 
	dotted line); adaptive mark I (+ and dash-dot line); and 
	adaptive mark II ($\times$ and dashed line).
	}
	\protect\label{fig5}
\end{figure}
	
\begin{figure}
	\caption{\narrowtext Plot of the exact  
	expressions for the log of the probability distribution $P_{\rm 
	coh}(\phi)$ for coherent states
	under the four schemes:  
	canonical (solid line); heterodyne (dotted line); 
	adaptive mark I (dash-dot line); and 
	adaptive mark II (dashed line). The coherent amplitude is 
	(a) $\beta = 1$, (b) $\beta=2$, (c) $\beta=3.5$, (d) $\beta=5$.
	}
	\protect\label{fig6}
\end{figure}

\begin{figure}
	\caption{\narrowtext Plot of the exact (points) and asymptotic (lines) 
	expressions for the log of the probability of error $E_{\rm coh}$ 
	for quaternary 
	phase encoding using coherent states of amplitude $\beta$ versus $\beta$ 
	under the four schemes:  
	canonical ($*$ and solid line); heterodyne ($\circ$ and 
	dotted line); adaptive mark I (+ and dash-dot line); and 
	adaptive mark II ($\times$ and dashed line).
	}
	\protect\label{fig7}
\end{figure}
 
\begin{figure}
	\caption{\narrowtext Plot of the exact (points)  
	expressions for the log of the minimum probability of error $E_{\rm coh}$ 
	for quaternary 
	phase encoding using the optimal state with at most $N$ photons versus $N$  
	under the four schemes:  
	canonical ($*$); heterodyne ($\circ$); adaptive mark I (+); and 
	}
	\protect\label{fig8}
\end{figure}

\end{multicols}

\begin{references}
\bibitem{fn2}
This result shows that we would only expect the asymptotic 
lower bound (\ref{minmin}) to be relevant only for $\ln N$ greater than $4 
\pi^{2}$; that is, for $N \agt 10^{18}$. Thus the absolute lower 
bound due to the intrinsic phase variance of the squeezed 
states which comprise the POM is irrelevant unless we have a photon 
detection scheme which is perfect to one part in $10^{18}$, 
which is practically impossible.
\end{references}
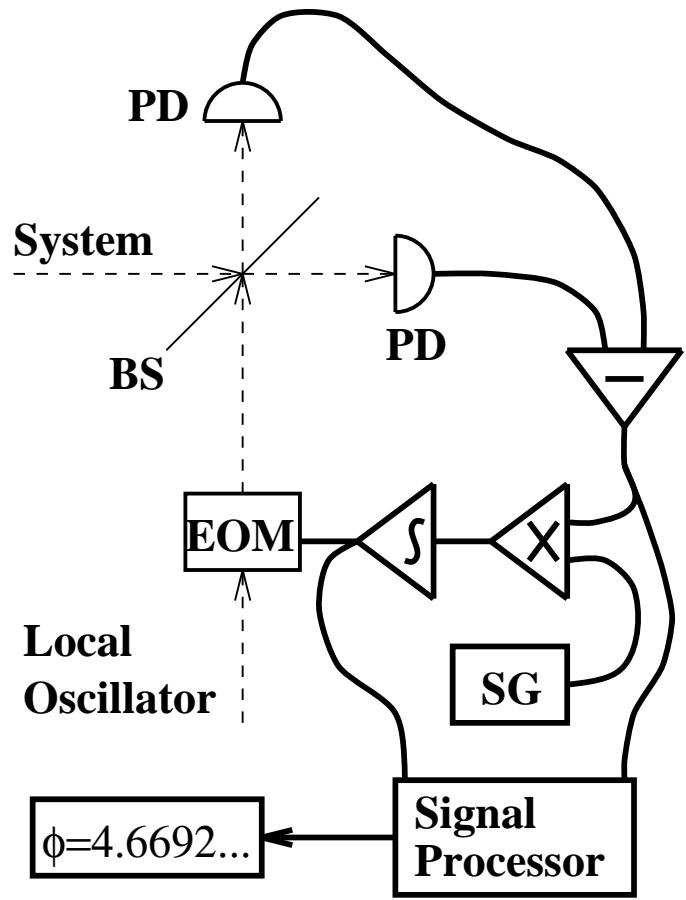
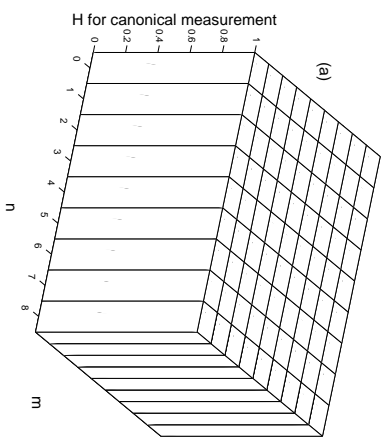
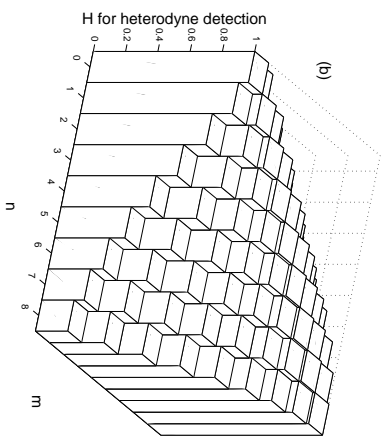
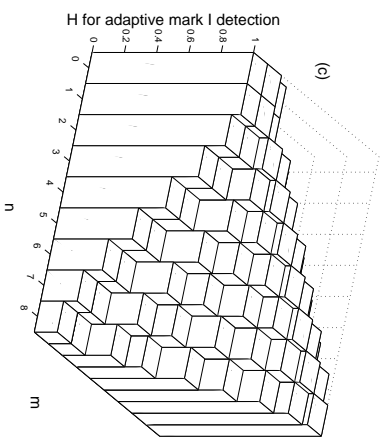
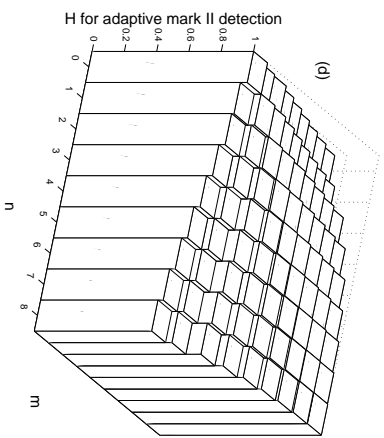
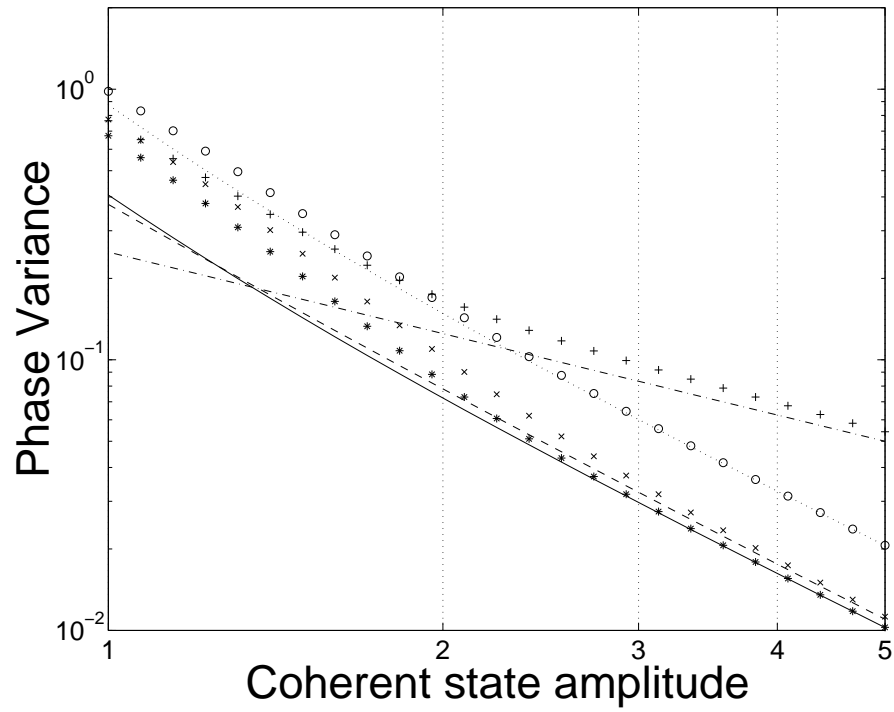
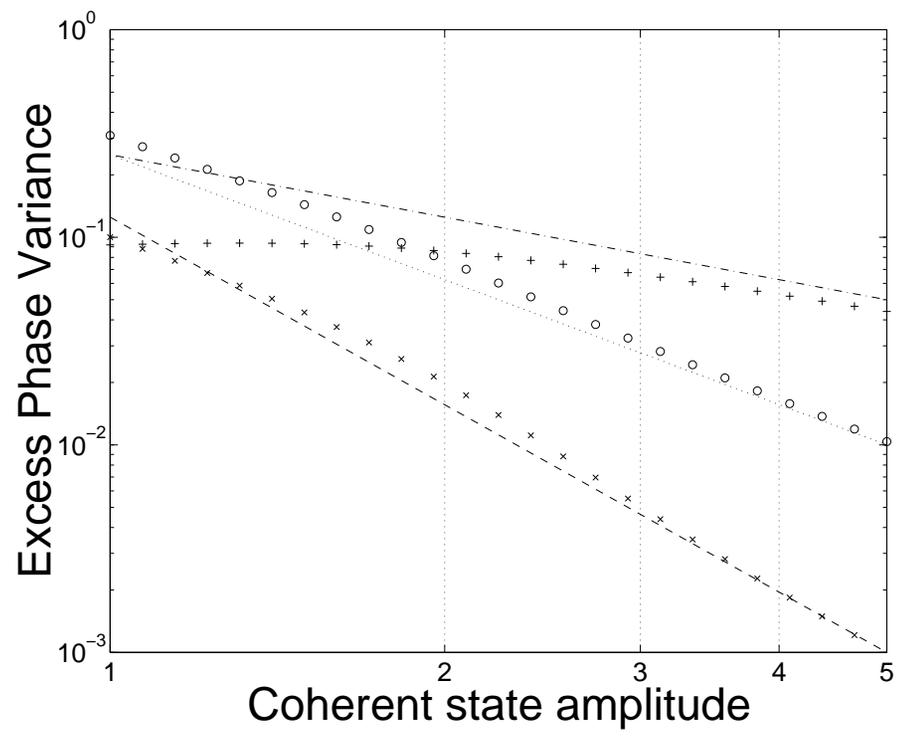
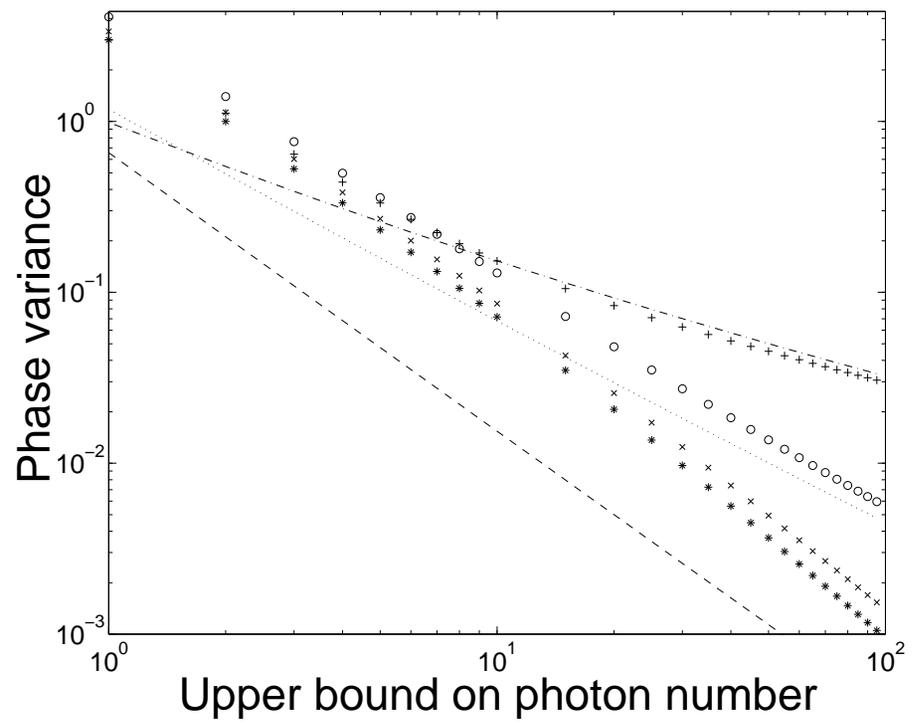
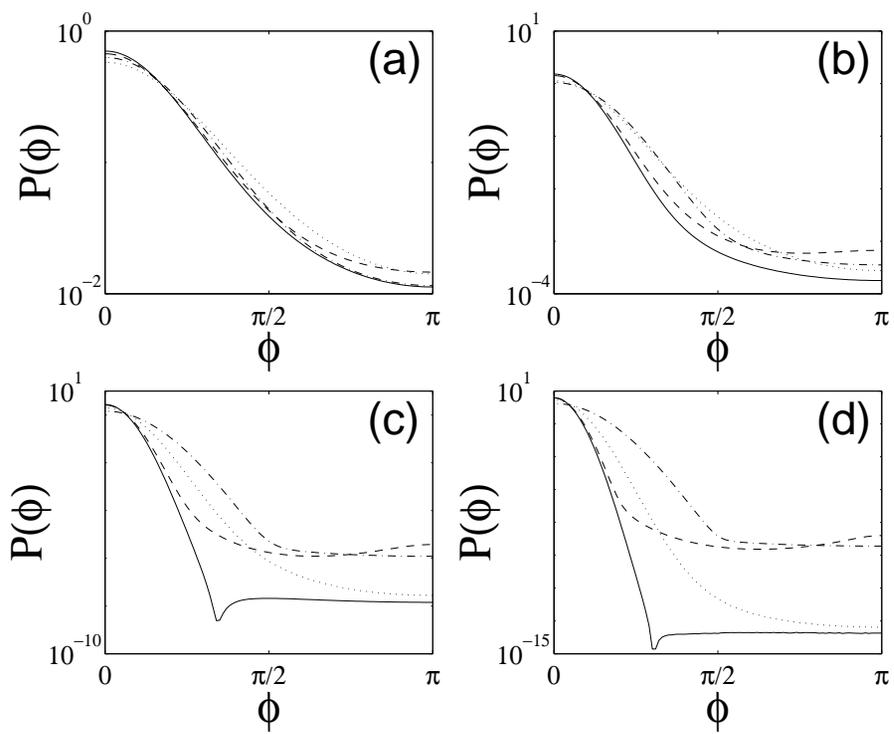
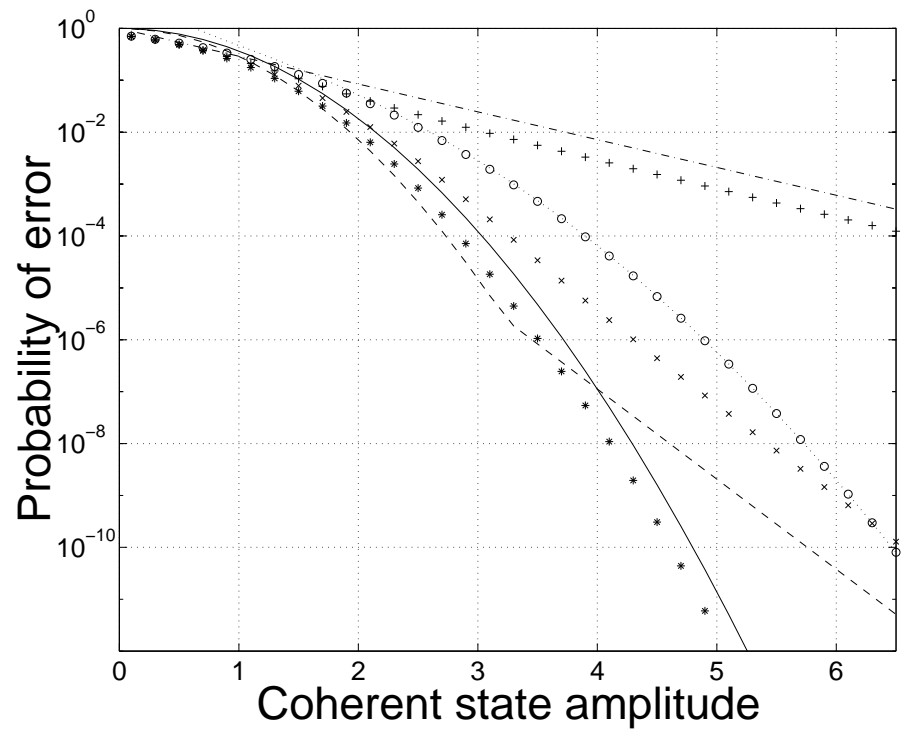
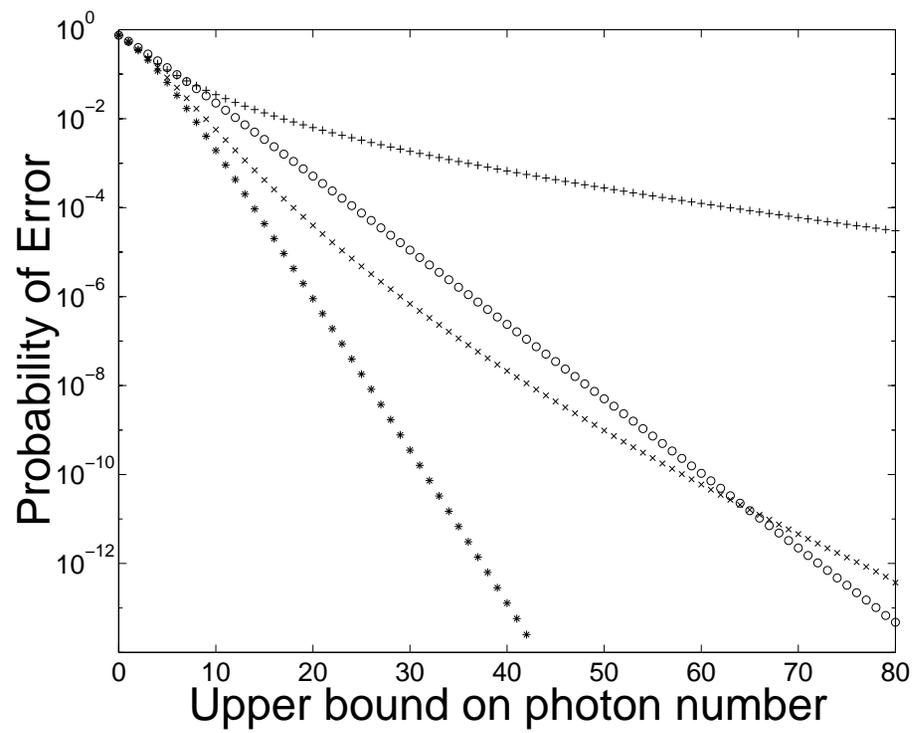
\end{document}